\documentclass[nofootinbib,11pt,superscriptaddress]{revtex4}
\pdfoutput=1
\usepackage[latin9]{inputenc}
\usepackage{color,graphicx,epsfig}
\usepackage{ifpdf}
\usepackage{amsmath}
\usepackage{bm}
\usepackage{color}
\usepackage[english]{babel}
\usepackage{graphicx}
\usepackage{amsfonts}
\usepackage{array}
\usepackage{multirow}
\usepackage{amssymb}
\usepackage{braket}
\usepackage{hyperref}

\definecolor{nicered}{rgb}{0.7,0.1,0.1}
\definecolor{nicegreen}{rgb}{0.1,0.5,0.1}
\hypersetup{colorlinks,citecolor= nicegreen,linkcolor= nicered}

\newcommand{\beq}{\begin{equation}}
\newcommand{\eeq}{\end{equation}}

\def\Cincy{Department of Physics, University of Cincinnati, Cincinnati, Ohio 45221,USA}
\def\FMF{Department of Physics,
  University of Ljubljana, Jadranska 19, 1000 Ljubljana, Slovenia}
\def\IJS{J. Stefan Institute, Jamova 39, P. O. Box 3000, 1001
  Ljubljana, Slovenia}
\def\LPSC{LPSC, Universit\'{e} Joseph Fourier Grenoble 1, CNRS/IN2P3 UMR5821, Institut Polytechnique de Grenoble, 53 rue des Martyrs, 38026 Grenoble Cedex, France}

\begin{document}

\title{Constraining Higgs mediated dark matter interactions}

\author{Admir Greljo}
\email[Electronic address:]{admir.greljo@ijs.si}
\affiliation{\IJS}

\author{J. Julio} 
\email[Electronic address:]{julio@ijs.si} 
\affiliation{\IJS}

\author{Jernej F. Kamenik} 
\email[Electronic address:]{jernej.kamenik@ijs.si} 
\affiliation{\IJS}
\affiliation{\FMF}

\author{Christopher Smith} 
\email[Electronic address:]{chsmith@lpsc.in2p3.fr} 
\affiliation{\LPSC}

\author{Jure Zupan} 
\email[Electronic address:]{zupanje@ucmail.uc.edu} 
\affiliation{\Cincy}

\begin{abstract}
We perform an analysis of Higgs portal models of dark matter (DM), where DM is light enough to contribute to invisible Higgs decays. Using effective field theory we show that DM can be a thermal relic only if there are additional light particles present with masses below a few 100 GeV. We give three concrete examples of viable Higgs portal models of light DM: (i) the SM extended by DM scalar along with an electroweak triplet and a singlet, (ii) a Two Higgs Doublet Model of type II with additional scalar DM, (iii) SM with DM and an extra scalar singlet that is lighter than DM. In all three examples the ${\mathcal B}(h\to {\rm invisible})$ constraint is not too restrictive, because it is  governed by different parameters than the relic abundance. Additional light particles can have implications for flavor violation and collider searches. 
\end{abstract}

\maketitle

\section{Introduction}

The narrow resonance with mass $m_h\simeq125$ GeV that was recently discovered at the LHC~\cite{Aad:2012tfa,Chatrchyan:2012ufa} is  a scalar \cite{key-1,ATLAS-CONF-2013-029,ATLAS-CONF-2013-031,Chatrchyan:2012jja,CMS-PAS-HIG-13-016} and has interactions consistent with those of the standard model (SM) Higgs boson \cite{key-2,{key-CMS}}. At present the experimental uncertainties are still relatively large and even ${\mathcal O}(1)$ deviations with respect to the SM couplings are possible. One of the more intriguing possibilities is that the Higgs could couple to dark matter (DM). 

The argument in favor of this possibility is quite general. Assuming that the discovered scalar is part of the Higgs electroweak doublet $H$, then  $H^\dagger H$ is the only gauge and Lorentz invariant  relevant operator in the SM. As such it can act as the ``Higgs portal" to DM~\cite{Patt:2006fw}.
The experimental searches place a number of nontrivial constraints on this idea. A pivotal parameter in the constraints is the DM mass. If DM is light, $m_{\rm DM}<m_h/2$, then Higgs can decay into DM. The resulting invisible decay width of the Higgs is bounded  at 95\% CL to $\mathcal B(h\to {\rm invisible})<0.19 (0.38)$ from global fits with the Higgs couplings to the SM fermions fixed to their SM values (varied freely while also allowing new particles in loops)~\cite{Belanger:2013xza} (see also~\cite{invHiggsBound}). This is a nontrivial constraint, since the SM Higgs decay width is so narrow. It essentially requires -- with some caveats to be discussed below -- that the Higgs coupling to DM needs to be smaller than roughly the SM bottom Yukawa coupling, $y_b\sim {\mathcal O}(0.02)$. This then insures that the invisible branching ratio is smaller than the dominant channel, $h\to b\bar b$.

On one hand we thus have a requirement that the Higgs should not couple too strongly to light DM. On the other hand, one needs ${\mathcal O}(1)$ couplings of Higgs to DM in order to obtain the correct thermal relic density. 
The tension between the two requirements leads to the apparent conclusion that the Higgs portal models with light DM are excluded. This was shown quantitatively in~\cite{Djouadi:2011aa} for the simplest models by assuming that  $\Gamma_h^{\rm invisible} \lesssim 0.2 \Gamma^{\rm SM}_h \simeq 0.8~\rm MeV$. Relaxing this bound by a factor of a few does not change the conclusion. 

For heavier DM, $m_{\rm DM}> m_h/2$, the bound on the invisible decay width of the Higgs is irrelevant. In this case one can search for DM using direct and indirect detection experiments. Existing constraints from direct DM detection are not stringent enough, but the next generation experiments are expected to cover most of the remaining viable parameter space~\cite{Aprile:2012zx}, with the exception of the parity violating Higgs portal where DM is a fermion~\cite{LopezHonorez:2012kv}. This, on the other hand, can be covered in the future using indirect DM searches~\cite{LopezHonorez:2012kv}. 

In this work we are primarily interested in the implications of an invisible Higgs decay signal (and the absence thereof so far) for light thermal relic DM. 
Are there still viable Higgs portal models with light DM? What modifications of the simplest models~\cite{Djouadi:2011aa} are needed? The conclusion that the simplest versions of the Higgs portal are excluded by the bound on $\mathcal B(h\to \rm invisible)$ utilizes effective field theory (EFT). The conclusion therefore relies on the assumption that an EFT description with the SM particles and DM as the only relevant dynamical degrees of freedom is valid both for the relic abundance calculation as well as for direct DM detection and Higgs phenomenology. 
For viable DM Higgs portals  then either the EFT description (with na\" ive power counting) must be violated, or the invisible decay width of the Higgs is naturally suppressed. As we will show below this implies that  {\em given present experimental constraints, the Higgs can couple significantly to thermal relic DM with mass less than half of the Higgs only if there are other light particles in the theory (barring fine-tuned situations).} In turn, should a nonzero invisible Higgs decay eventually be found and interpreted as a decay to thermal relic DM particles, then other new light particles need to be discovered. 

To demonstrate this we first show in Section~\ref{sec:2} that extending the EFT description to higher dimensional operators but not enlarging the field content does not change the conclusions about  the minimal DM Higgs portals if $h\to$DM+DM decay is allowed. In Section~\ref{sec:suppressed-decays} we then show that for models where the two body Higgs decays to dark sector are forbidden, the scale of the EFT is small, $\Lambda\sim { \mathcal O}({\rm few~100~GeV})$. This again implies that viable Higgs portals of DM require new light degrees of freedom beyond SM+DM. In Section~\ref{sec:4} we in turn give three examples of viable Higgs portal models of DM. Two models, described in subsections~\ref{subsec:A} and~\ref{thdm}, can be matched onto EFT since the additional degrees of freedom are heavier -- though not much heavier -- than the Higgs. The two models do require fine-tuned cancellations in order to avoid experimental constraints.  A model discussed in subsection~\ref{subsection:C}, on the other hand, requires no such tunings. It contains, however, a particle lighter than DM and therefore violates the EFT assumptions. 
We summarize our conclusions in Section \ref{sec:conclusions}. Details on direct DM detection, relic abundance calculations, and the fits to the Higgs data are relegated to the Appendices \ref{sec:3} and \ref{app:Higgs:fit}, respectively.

\section{Higgs portals in Effective Field Theory}
\label{sec:2}

We start by reviewing the minimal Higgs portal scenarios. The SM is enlarged by a single neutral (DM) field, odd under a $Z_2$ symmetry. In the following we consider DM with spins up to and including spin $1$, i.e. the possibility that DM is a scalar, $\phi$, a fermion, $\psi$ or a vector, $V_\mu$. The dominant interactions of  DM with the SM are in each case, respectively,
\begin{subequations}
\begin{align}
\label{eq:op0}
\mathcal H^0_{\rm eff} &= \lambda' H^\dagger H \times \phi^\dagger\phi\,, \\
\label{eq:op12}
\mathcal H^{1/2}_{\rm eff} &= \frac{c_{S}}{\Lambda} H^\dagger H \times \bar \psi \psi + \frac{i c_{P}}{\Lambda} H^\dagger H \times \bar \psi \gamma_5 \psi \,, \\
\mathcal H^{1}_{\rm eff} &= \epsilon_H H^\dagger H \times V^\mu V_\mu\,.
\label{eq:op32}
\end{align}
\end{subequations}
After electroweak (EW) symmetry breaking
\begin{equation}
H^\dagger H \to \frac{1}{2} (v_{\rm EW}^2 + 2 v_{\rm EW} h + h^2)\,,
\end{equation}
 where $v_{\rm EW}\simeq 246$~GeV is the electroweak condensate and $h$ the Higgs boson. We see that the scalar and vector DM have renormalizable Higgs portal interactions with the SM. For fermion DM these interactions start only at dimension 5. In Eq. \eqref{eq:op12} $\Lambda$ is the scale at which the non-renormalizable DM-Higgs interactions are generated. In principle one can also write down higher dimensional operators that supplement \eqref{eq:op0}-\eqref{eq:op32}, but are suppressed by more powers of $\Lambda$. The minimal Higgs portal models of DM assume $\Lambda\gg v_{\rm EW}, m_{\rm DM}$, such that the expansion $v_{\rm EW}/\Lambda$ makes sense and \eqref{eq:op0}-\eqref{eq:op32} are the dominant contributions to DM-SM interactions in the early universe and current experiments. As shown in~\cite{Djouadi:2011aa}, in all such models with light DM ($m_{\rm DM}\lesssim m_h/2$), the observed DM relic abundance is in conflict with the experimental bounds on the invisible decay width of the Higgs, while in the region $m_{\rm DM} > m_h/2$, direct DM detection experiments are beginning to exclude the remaining parameter space.

But would the situation change if the $v_{\rm EW}/\Lambda$ expansion would not start at the lowest order, Eqs. \eqref{eq:op0}-\eqref{eq:op32}? Can higher dimensional Higgs-DM operators~\cite{Kamenik:2012hn} open new possibilities to reconcile Higgs portal DM with current experimental constraints?
To answer this question we first perform a na\"ive dimensional analysis of the relevant processes based solely on the canonical dimension ($d=4+n$) of the relevant interaction operator. For $m_{\rm DM} \ll m_h/2$ the invisible Higgs branching fraction scales as
\begin{equation}
\mathcal B( h\to{\rm invisible}) \sim  10^{3} \left( \frac{m_h}{\Lambda} \right)^{2n}\,,\label{eq:invisible:scale}
\end{equation}
where the overall normalization, $10^3\sim 1/y_b^2$, is set by the total width of the SM Higgs.
In \eqref{eq:invisible:scale} we used $v_{\rm EW}\sim m_h$, assumed that all dimensionless DM--Higgs couplings are ${\mathcal O}(1)$,  and also assumed two-body  $h\to{\rm invisible}$ decay kinematics. In comparison, the current constraints from direct DM detection experiments give 
\begin{equation}
\frac{\langle \sigma_{\rm dir} \rangle}{\langle \sigma_{\rm dir}\rangle_{\rm excl.}} \sim 10^2  \left( \frac{m_h}{\Lambda} \right)^{2n}\left(\frac{m_{\rm DM}}{m_h}\right)^{m} \beta^{2m'}\,,\label{eq:dir:det}
\end{equation}
where $m,m'$ are non-negative integers, while the numerical pre-factor is simply the translation of the experimental limit due to XENON100~\cite{aprile:2012} and will increase in the future. Note that \eqref{eq:dir:det} assumes spin independent scattering since this is stronger than spin dependent one. The suppression in terms of $m_h/\Lambda$ is the same as for ${\mathcal B}(h\to {\rm invisible})$, but depending on the operator structure there may be additional suppressions from typical DM velocity in the galactic halo, $\beta \sim 10^{-3}$, or from DM mass insertions, $m_{\rm DM}/m_h$. Both of these factors are smaller than one, therefore we conclude that at present for light DM  the Higgs constraints are stronger than direct DM detection constraints for any operator dimension. 

If DM is a thermal relic, then its abundance is fixed by thermal DM annihilation cross-section at the time of freeze-out,
\begin{equation}
{\langle \sigma_{\rm ann.} v \rangle} \propto  \frac{y_f^2}{m_h^2} \left( \frac{m_{h} }{\Lambda} \right)^{2n}  \left( \frac{m_{\rm DM} }{m_h} \right)^{k} \,,\label{eq:annih:scaling}
\end{equation}
where $y_f$ is the SM Yukawa coupling for the heaviest open SM fermion channel, and $k\geqslant k_{\rm min}=0 (2)$ for scalar and vector (fermion) DM with the equality sign for the lowest dimensional operators. In \eqref{eq:annih:scaling} we neglected relative velocity suppressions, $v_r\sim 0.4$, and as before set all Wilson coefficients to be ${\mathcal O}(1)$. In order to obtain the correct relic density, ${\langle \sigma_{\rm ann.} v \rangle}\simeq 3 \cdot 10^{-26} {\rm cm}^3/{\rm s}$, with $\Omega_{\rm DM}\propto 1/\langle \sigma_{\rm ann.} v \rangle$. From Eq. \eqref{eq:annih:scaling} we then see that  the correct relic density requires the scale $\Lambda$ to be lower if the dimensionality $n$ of the operator  setting the annihilation cross section is higher. The scaling of ${\mathcal Br}(h\to {\rm invisible})$ in terms of $\Lambda$ is the same as for ${\langle \sigma_{\rm ann.} v \rangle}$, so that for the correct relic density one has
\begin{equation}
\left(  \frac{\mathcal B_h^{\rm invis.}}{\langle \sigma_{\rm ann.} v \rangle^{\rm }} \right)_n \sim
\left(\frac{m_h}{m_{\rm DM}}\right)^{k-k_{\rm min}} \left(  \frac{\mathcal B_h^{\rm invis.}}{\langle \sigma_{\rm ann.} v \rangle^{\rm }} \right)_{n_{\rm min}},
\end{equation}
where $n_{\rm min}=4(5)$ for scalar and vector (fermion) DM. Since $k-k_{\rm min}\geqslant 0$, the Higgs constraints can only become stronger if the Higgs portal proceeds through higher dimensional operators. As a result, the higher dimensional operators cannot reconcile Higgs portal DM with the bounds on invisible Higgs branching ratio as long as $h\to {\rm DM} +{\rm DM}$ is possible and all couplings are ${\mathcal O}(1)$.

\section{Suppressed Higgs decays to dark sector}
\label{sec:suppressed-decays}

In the previous section we saw that $\mathcal B(h\to {\rm invisible})$ places strong constraints on Higgs portals of DM. The analysis relied on two assumptions, {\rm i)} that $h\to$DM+DM decay is possible, and {\rm ii)} that DM is the only light new physics particle. In this section we investigate in more details the first assumption, while the second assumption will be relaxed in the subsequent section. In the remainder of this section we therefore assume that $h\to$DM+DM decay is forbidden either accidentally or due to the structure of the theory.

There are three possibilities to suppress the $h\to$DM+DM decay. The first one is to assume DM annihilation to SM particles proceeds predominantly through operators not involving the Higgs. This possibility is orthogonal to the basic idea of a Higgs portal. It has also been studied extensively (c.f.~\cite{dmref}) and we do not pursue it any further. The second possibility is that the $h\to$DM+DM decay is kinematically forbidden simply because DM is heavy enough, $m_{\rm DM}>m_h/2$. The final possibility is that DM couples through a special subset of Higgs portal operators, such that $h\to$DM+DM decay is forbidden, while $h\to$DM+DM+$X_{\rm SM}$ is allowed, where $X_{\rm SM}$ denotes one or more SM particles in the final state. We set aside the model building question of how this is arranged in the UV theory and work within EFT. The $y_b^2$ suppression of the SM Higgs decay width is roughly of the same size as the phase space suppression from one or two additional final state particles.  One may thus expect that ${\mathcal O}(1)$ couplings between DM and the Higgs would give at the same time the correct relic abundance as well as small enough ${\mathcal B}(h\to 2{\rm DM}+X)$. Below we go through a list of possible operators, and as we will see a number of them are not excluded by direct and indirect DM detection constraints.

\begin{figure}
\centering
\includegraphics[scale=0.8]{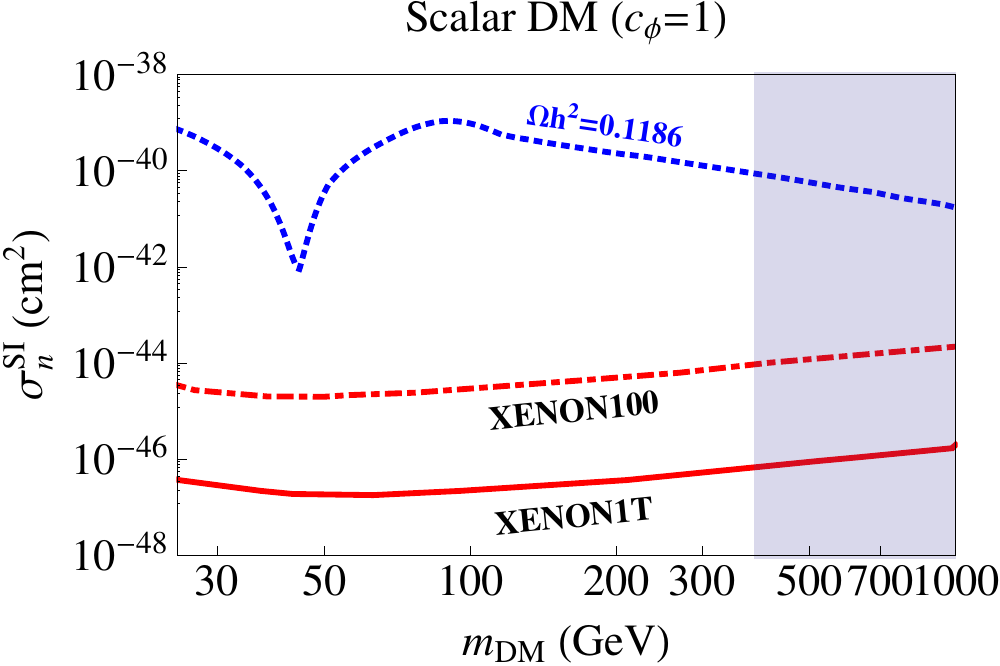}
\includegraphics[scale=0.8]{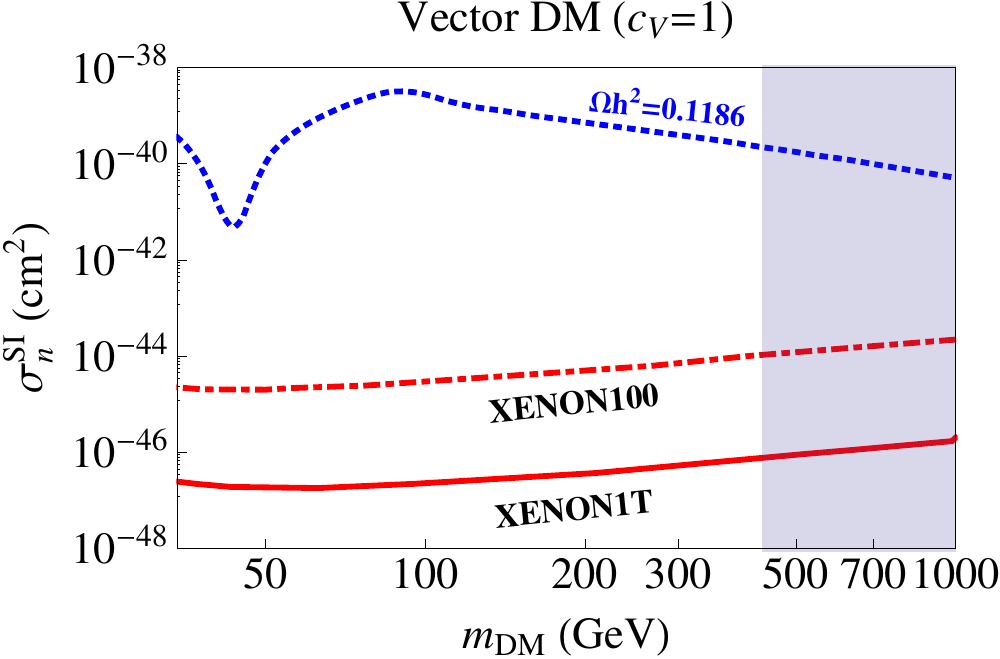}

~

\includegraphics[scale=0.8]{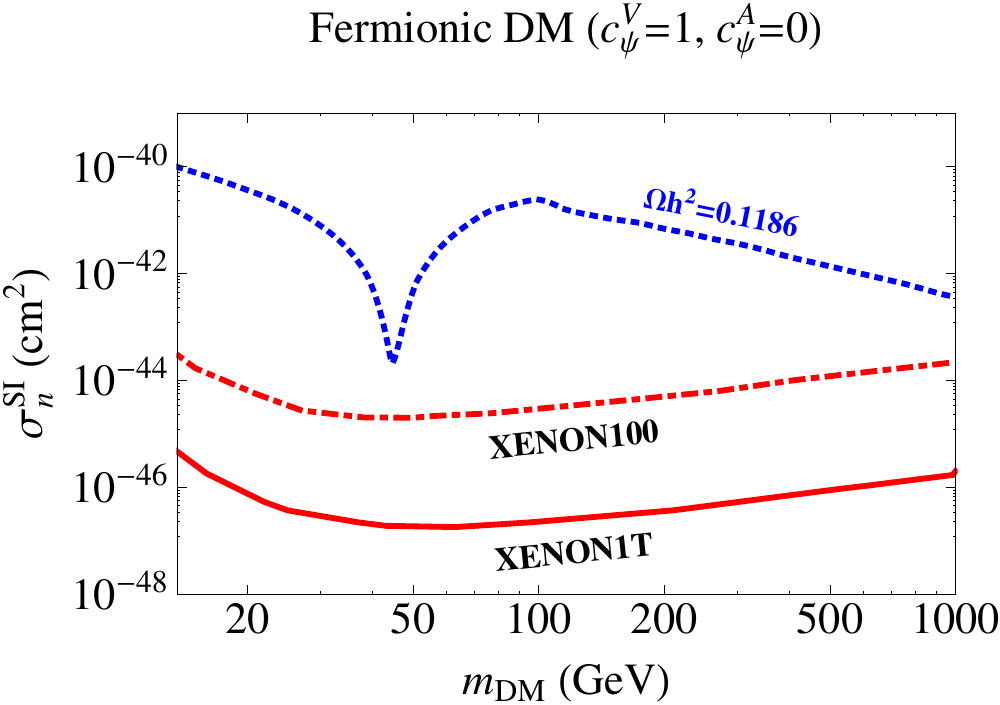}
\includegraphics[scale=0.8]{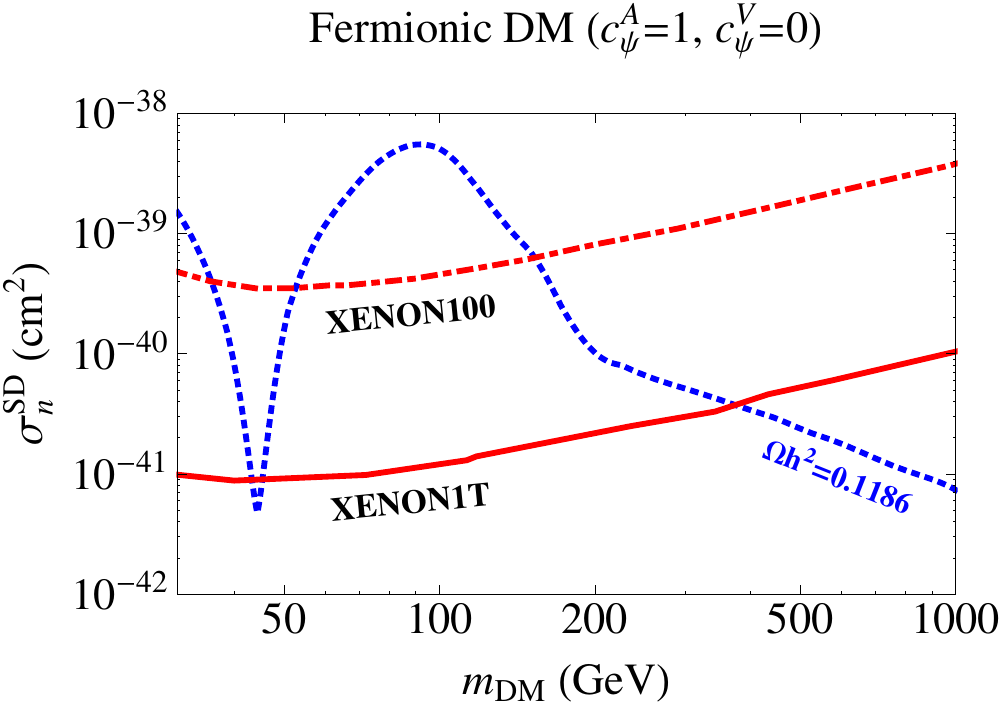}
\caption{The spin independent DM-nucleon cross sections (dashed-blue) induced by Higgs vector current operators (\ref{GaugeOp}) after requiring correct thermal relic density $\Omega_{\rm DM}h^2=0.1186\pm 0.0031$ \cite{Ade:2013ktc} for scalar DM (top left), vector DM (top right) and fermion DM with vector (bottom left) couplings. Bottom right panel shows the spin dependent cross section for fermion DM with axial vector couplings. The current XENON100 \cite{aprile:2012} and projected future XENON1T bounds \cite{Aprile:2012zx} are denoted by dot-dashed and solid red lines, respectively. The shaded blue regions indicate where the EFT description breaks down ($\Lambda < 2 m_{\rm DM}$ ).}
\label{gaugech_dd}
\end{figure}

The simplest effective interactions generating $h\to{\rm DM}+{\rm DM}+X_{\rm SM}$ decays are built from the Higgs vector current 
\begin{equation}
H^{\dagger}\overleftrightarrow{D}\hspace{0in}^{\mu}H\equiv H^{\dagger
}\overleftarrow{D}\hspace{0in}^{\mu}H-H^{\dagger}\overrightarrow{D}%
\hspace{0in}^{\mu}H\rightarrow\frac{ig}{2c_{W}}(v_{\rm EW}^{2}+2v_{\rm EW}h^{}+h^{2}%
)Z^{\mu}\;,\label{HDH}%
\end{equation}
where $c_W = \cos \theta_W$, with $\theta_W$ the weak mixing angle. The operators of the lowest dimension are~\cite{Kamenik:2012hn} 
\begin{subequations}
\label{GaugeOp}%
\begin{align}
\mathcal{H}_{\rm eff}^{0} &  =\frac{c_{\phi}}{\Lambda^{2}}H^{\dagger
}\overleftrightarrow{D}\hspace{0in}_{\mu}H\times\phi^{\dagger}%
\overleftrightarrow{\partial}\hspace{0in}^{\mu}\phi\;,\label{GaugeOp1}\\
\mathcal{H}_{\rm eff}^{1/2} &  =\frac{c_\psi^V}{\Lambda^{2}}iH^{\dagger
}\overleftrightarrow{D}\hspace{0in}_{\mu}H\times\bar{\psi}\gamma^{\mu}%
\psi + \frac{c_\psi^A}{\Lambda^{2}}iH^{\dagger}\overleftrightarrow
{D}\hspace{0in}_{\mu}H\times\bar{\psi}\gamma^{\mu}\gamma_5\psi%
\;,\label{GaugeOp2}\\
\mathcal{H}_{\rm eff}^{1} &  = \frac{c_V}{\Lambda^2}iH^{\dagger}\overleftrightarrow
{D}\hspace{0in}_{\nu}H  \times V_{\mu}\overleftrightarrow{\partial}^\nu V^{\mu}\,.
 \;\label{GaugeOp3}
\end{align}
\end{subequations}
For example, they appear in models where the DM is charged under a hidden $U(1)$ gauge symmetry (spontaneously broken above the weak scale), exhibiting kinetic mixing with the SM hypercharge~\cite{Holdom:1985ag}.
These operators induce a three body   decay $h\to{\rm DM}+{\rm DM}+Z$, where for $Z\to \nu\bar\nu$ the decay would be completely invisible. They do not lead, however, to two body invisible  decay $h\to{\rm DM}+{\rm DM}$. The three body Higgs decay is kinematically allowed if $m_{\rm DM} < (m_h-m_Z)/2\simeq 17$~GeV. Such a light DM is subject to  bounds from $Z\to E_{\rm miss}$ 
measurements at LEP~\cite{pdg}. Requiring the correct relic density this constrains $m_{\rm DM}>24(34)$~GeV for scalar (vector) DM, and $m_{\rm DM}>14(31)$~GeV for fermionic DM with vector (axial-vector) interaction. 

The operators in Eq.~(\ref{GaugeOp}) are also subject to severe direct DM detection constraints from $Z$-mediated DM scattering on nuclei (for details see Appendix \ref{sec:3}).
In Fig.~\ref{gaugech_dd} we show the predicted spin independent DM-nucleon cross sections (dashed blue lines) after requiring the correct thermal relic density $\Omega_{\rm DM}h^2=0.1186\pm 0.0031$ \cite{Ade:2013ktc}.  The shaded blue regions indicate the validity of EFT, i.e., that $\Lambda \geq 2m_{\rm DM}$. With the exception of fermionic DM with purely axial-vector interaction ($c_\psi^V=0$)  all parameter space allowed by relic density is excluded by XENON100  \cite{aprile:2012} (dot-dashed red lines). For fermionic DM with purely axial-vector interactions the spin-dependent cross section is plotted in Fig.~\ref{gaugech_dd}, bottom right panel, since the SI cross-section is velocity suppressed. The result is compared to recent XENON100 bound on SD DM-neutron cross section~\cite{aprile:2013},
which excludes $m_{\rm DM}< 35$~GeV and 50~GeV$<m_{\rm DM}<$150~GeV. Note that the XENON1T~\cite{Garny:2012it} is expected to cover almost completely the remaining low DM mass window.  In summary, the combination of invisible $Z$ decay and direct DM detection constraints excludes any appreciable ${\mathcal B}(h\to {\rm invisible})$ from operators in Eq. \eqref{GaugeOp}. 

Another possibility is to couple DM to scalar or tensor fermionic currents. These 
automatically involve a Higgs
field,
\begin{equation}
\Gamma^{S}=H^{\dagger}\bar{D}Q,\;\;H^{\dagger}\bar{E}L,\;\;H^{\ast\dagger
}\bar{U}Q,\;\;\;\;\Gamma_{\mu\nu}^{T}=H^{\dagger}\bar{D}\sigma_{\mu\nu
}Q,\;\;H^{\dagger}\bar{E}\sigma_{\mu\nu}L,\;\;H^{\ast\dagger}\bar{U}\sigma
_{\mu\nu}Q\;.\label{eq:currents}
\end{equation}
The lowest dimensional operators are then
\begin{subequations}
\label{HFermion}%
\begin{align}
\mathcal{H}_{\rm eff}^{0} &  =\frac{f_\phi}{\Lambda^{2}}\Gamma^{S}\times\phi^{\dagger}%
\phi\; +h.c.,\\
\mathcal{H}_{\rm eff}^{1/2} &  =\frac{f_{\psi}^S}{\Lambda^{3}}\Gamma^{S}%
\times\overline{\psi}\psi + \frac{f_{\psi}^P}{\Lambda^{3}}\Gamma^{S}%
\times i\overline{\psi}\gamma_5\psi + \frac{f^{T}_\psi}{\Lambda^{3}}\Gamma_{\mu\nu}%
^{T}\times\overline{\psi}\sigma^{\mu\nu}\psi\; +h.c.,\\
\mathcal{H}_{\rm eff}^{1} &  =\frac{f_V}{\Lambda^{2}}\Gamma^{S}\times
V_{\mu} V^{\mu}\;+h.c.,
\end{align}
\end{subequations}
where the dependence of couplings on SM fermion flavors is implicit. Operators involving $\Gamma^S$ can be generated for example in models with extended scalar sectors, as we will discuss below. On the other hand, the generation of tensorial $\Gamma^T_{\mu\nu}$ interactions is typically more involved. One possibility is to introduce a SM-DM mediator sector with a gauge symmetry under which both SM and DM are neutral. The appropriate irrelevant couplings to generate the tensorial SM-DM interaction can then possibly be obtained at the loop level. A complete model construction is thus quite intricate and beyond our scope, so we do not pursue it any further.

We first assume  the couplings in Eqs. (\ref{HFermion}) to be proportional to the fermion masses,
\beq
f_\phi = \frac{\sqrt{2}m_f}{v_{EW}},~~ f_{\psi}^{S,P,T} = \frac{\sqrt{2}m_f}{v_{EW}},~~f_V=\frac{\sqrt{2}m_f}{v_{EW}},
\label{cassumpt}
\eeq
so that possible flavor changing neutral currents (FCNCs) are automatically suppressed. The operators in Eq.~\eqref{HFermion} lead to four body Higgs decays, that are unobservably small. For instance, assuming thermal relic DM with $m_{\rm DM}=20$ GeV one has ${\cal B}(h \to {\rm DM+DM}+b\bar{b})\sim {\mathcal O}(10^{-7})$ for both purely pseudo scalar and purely tensorial DM interactions. 

\begin{figure}
\centering
\includegraphics[scale=0.8]{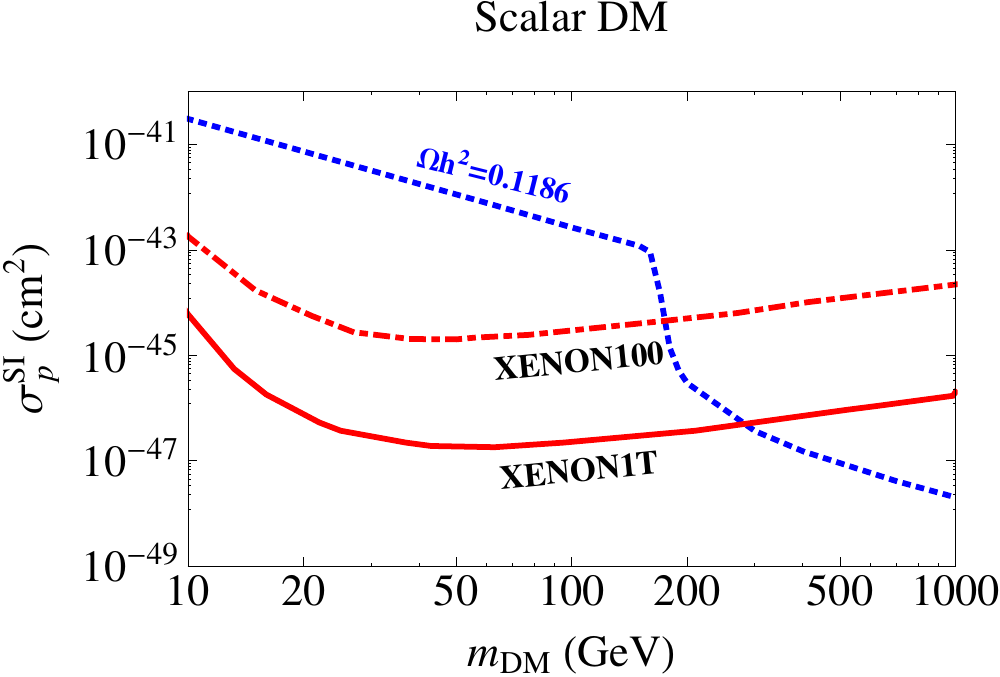}
\includegraphics[scale=0.8]{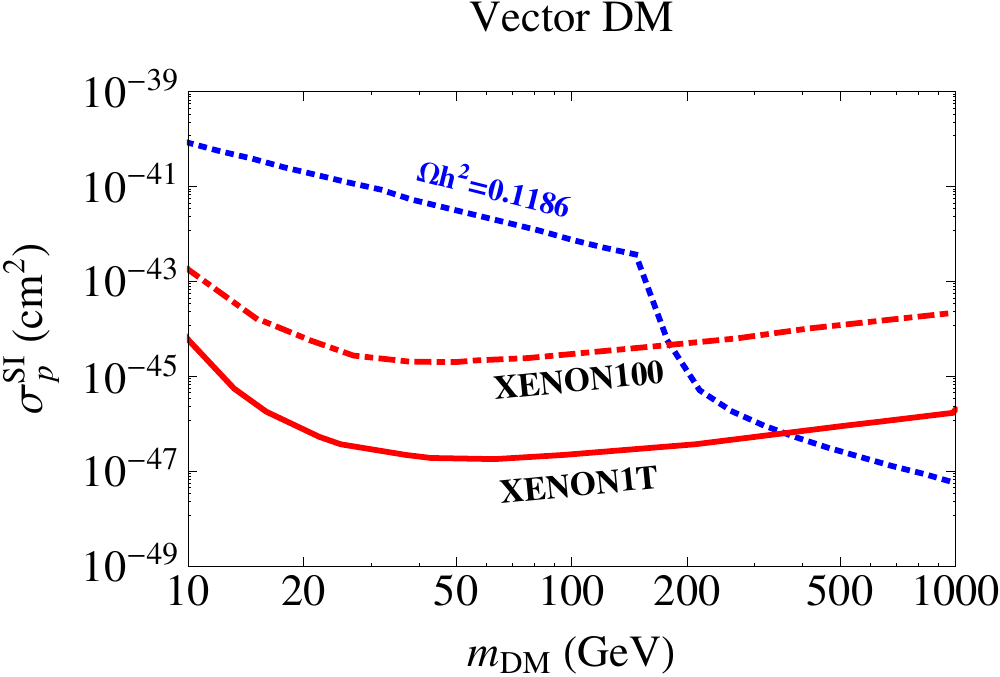}
\includegraphics[scale=0.8]{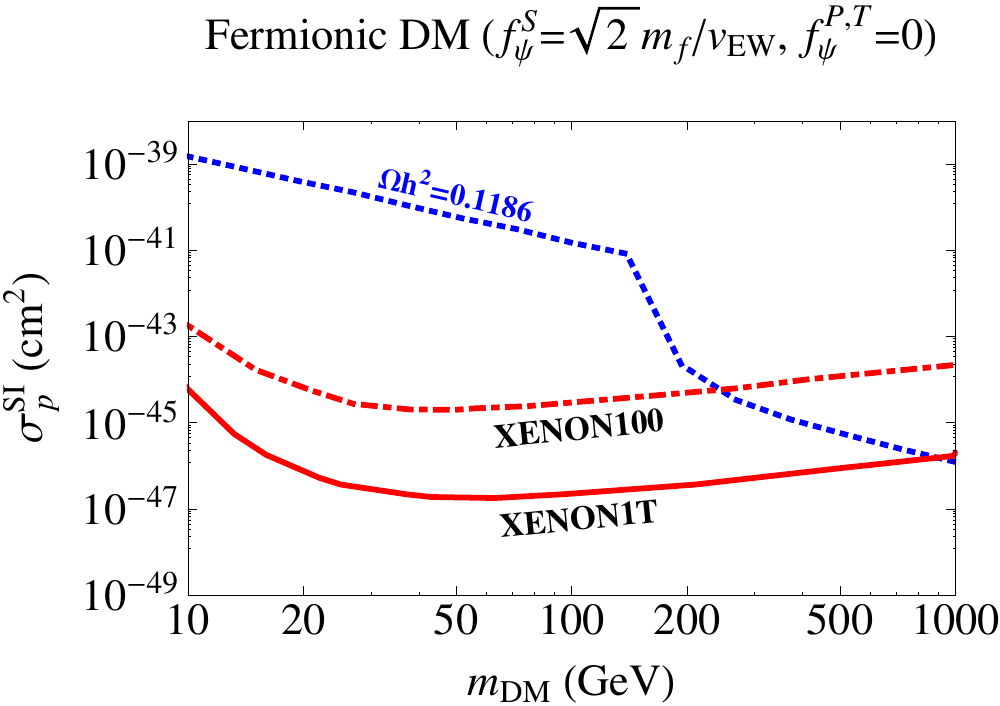}
\includegraphics[scale=0.8]{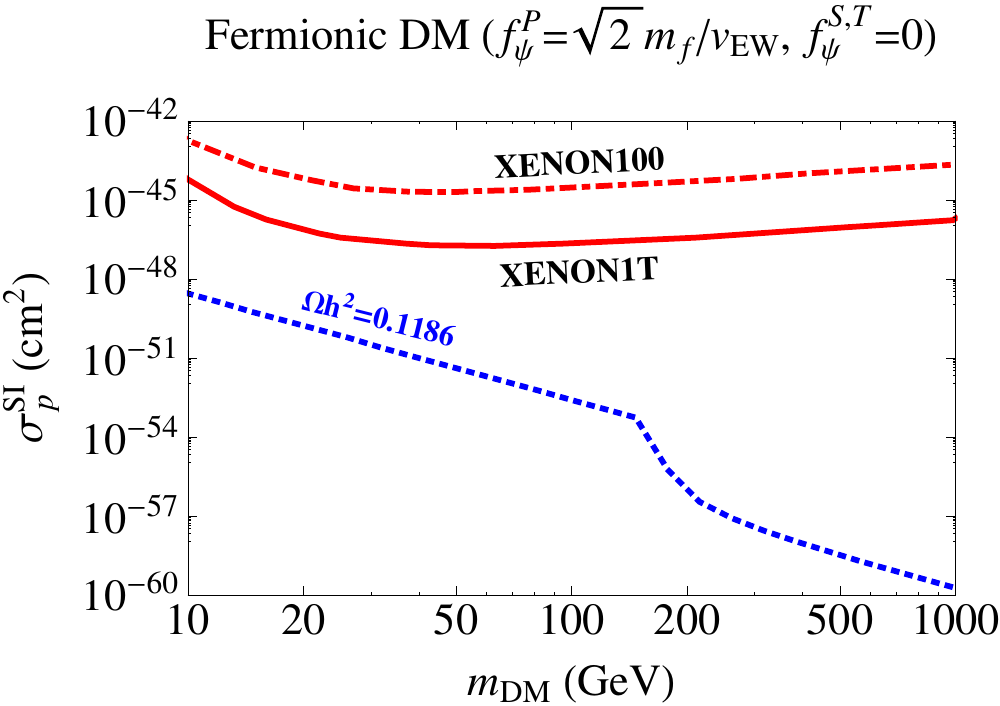}
\includegraphics[scale=0.8]{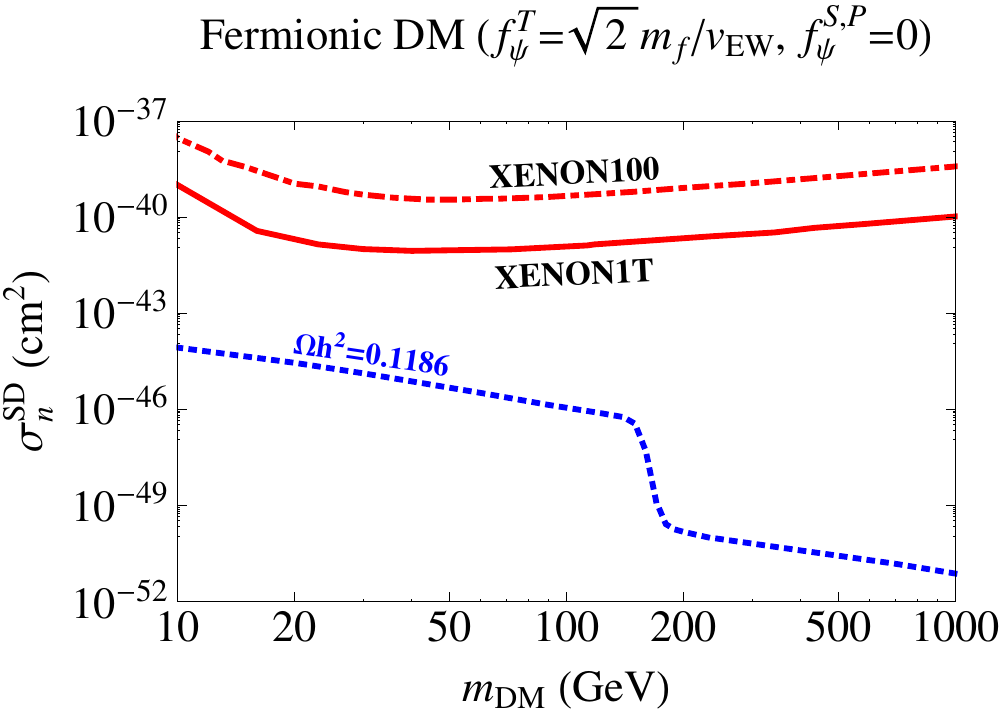}
\caption{The DM-nucleon cross sections (dashed-blue) induced by operators (\ref{HFermion}). The predicted values are compared to the current XENON100 bound (dot-dashed-red line) and future XENON1T bound (continuous-red line).
}
\label{fermionch_dd}
\end{figure}

Fig.~\ref{fermionch_dd} shows the  predictions for the spin-independent DM-nucleon cross sections in the upper four panels, for scalar DM, vector DM, and fermion DM with scalar and pseudoscalar interactions, respectively (blue dashed lines), requiring correct thermal relic DM abundance. The spin-dependent cross section for fermion DM with purely tensorial interaction is shown in the lower panel in Fig. \ref{fermionch_dd}.
For the chosen flavor structure of the relevant couplings, Eq.~\eqref{cassumpt},  XENON100 bounds (dot-dashed red lines) exclude almost all possibilities except for fermionic DM with parity-violating or tensorial interactions. The parity violating fermionic DM evades the current XENON100 and also the projected XENON1T bound (red solid line) because the scattering cross section is velocity suppressed. The direct detection cross section for the tensorial interactions is strongly suppressed by the  assumption that the coupling to light quarks is suppressed by light quark masses, Eq. (\ref{cassumpt}) (unlike for scalar interactions this suppression carries over for tensor interactions when matching from quark to nucleon level operators, see Refs.~\cite{belanger} and~\cite {fan10} for further details).

\begin{figure}[t!]
\centering
\includegraphics[scale=0.8]{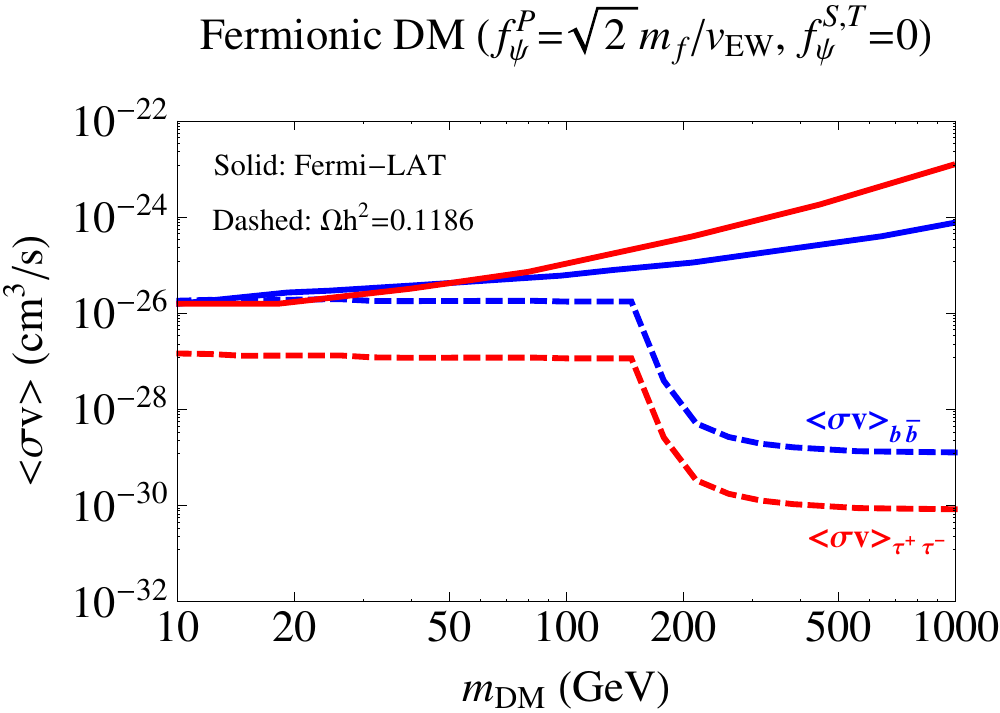}
\includegraphics[scale=0.8]{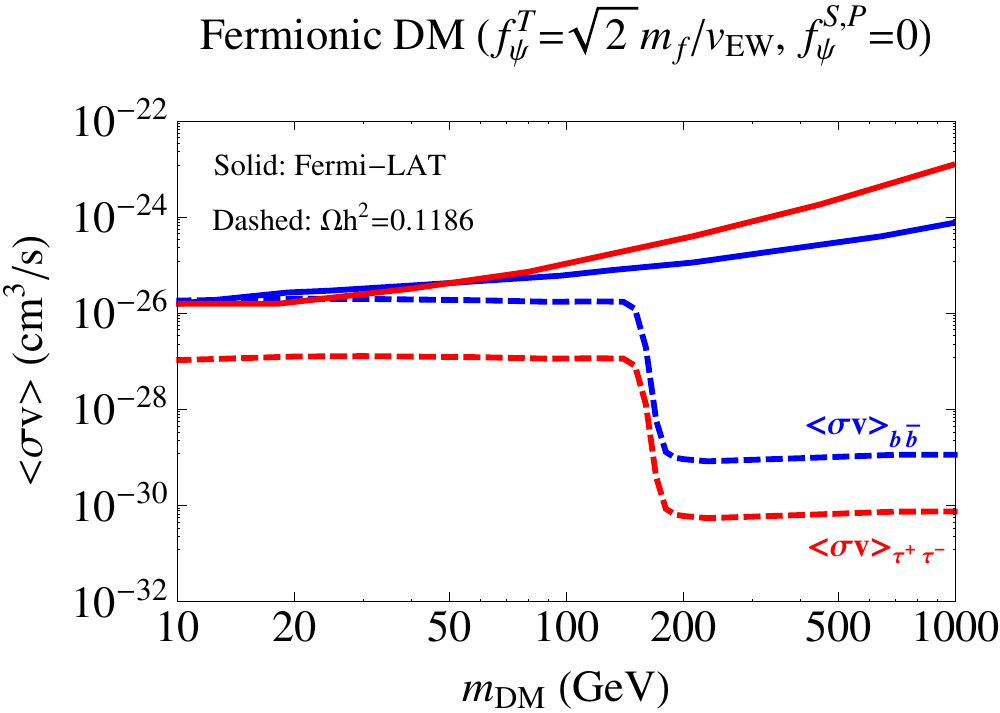}
\includegraphics[scale=0.8]{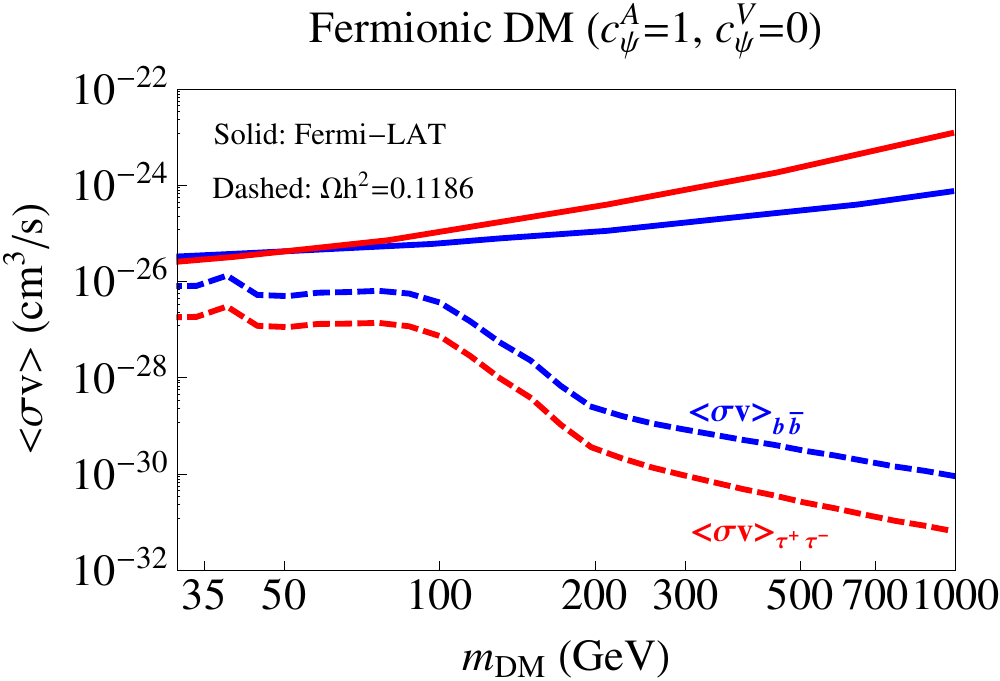}
\caption{The $b\bar{b}$ (blue) and $\tau^+\tau^-$ (red) annihilation cross-sections ($\left<\sigma v\right>$) for the fermionic operators in \eqref{HFermion} (upper two panels) and for fermionic DM with axial vector coupling to Higgs vector current in \eqref{GaugeOp} ($c_\psi^V=0$). The continuous (dashed) lines indicate the present experimental upper bounds~\cite{dsph-fermi,garinger-sameth} (predicted values assuming correct DM relic density) on $\left<\sigma v\right>$.
}
\label{id-bound}
\end{figure}

The remaining two possibilities are constrained by indirect DM searches. In Fig.~\ref{id-bound} we compare the bounds on annihilation cross sections 
$\left<\sigma v \right>$ for $b\bar{b}$ (blue lines) and $\tau^+\tau^-$ (red lines) channels \cite{dsph-fermi,garinger-sameth} with the predictions from the last two operators in Eq.~\eqref{HFermion}, when correct relic density is assumed in the predictions. We see that the fermionic DM with pseudo-scalar or tensorial interactions is constrained to be heavier than $m_{\rm DM}> 15$~GeV. For reference we also show in Fig.~\ref{id-bound} the possibility of Higgs portal coupling to DM through the axial-vector operator from Eq.  \eqref{GaugeOp}, which is not excluded by direct detection. It demonstrates that for $Z$ mediated channels, the constraints from indirect detection are not as significant. The reason lies in the assumed flavor structure. This is fixed for operators in Eq.~\eqref{GaugeOp} by the couplings of the $Z$. DM then annihilates to all fermions democratically, reducing the signal in the $b\bar b$ and $\tau^+ \tau^-$ final states. For the flavor structure assumed in Eq.~\eqref{HFermion} these are the dominant channels, however, making the constraints more powerful. 

This also highlights the fact that the bounds on operators in Eq.~\eqref{HFermion} depend strongly on the assumed flavor structure of the Wilson coefficients. We do not attempt to cover all possibilities but rather only entertain a few representative cases. For instance,  increasing the couplings to light quarks, $u,d,s$, the  direct DM detection bounds would become significantly stronger, while the relic density would remain practically unaffected. Note that in the limit where DM does not couple to the light quarks but only to $3^{\rm rd}$ generation, the direct detection bounds are still relevant since one induces interactions to gluons at loop level.
An interesting possibility is to have Wilson coefficients differ in sign such that the DM-nucleon elastic scattering cross-section vanishes. This possibility was pointed out in the context of type II Two-Higgs-Doublet Model (2HDM-II) in Ref. \cite{He:2008qm}, to be discussed in more detail in subsection~\ref{thdm}. Another possibility where direct detection bounds are weak or completely irrelevant is the case of leptophilic DM, where the Wilson coefficients for operators coupling to quarks in Eq.~\eqref{HFermion}  are suppressed~\cite{Kopp:2010su}. 

\begin{figure}
\centering
\includegraphics[scale=1]{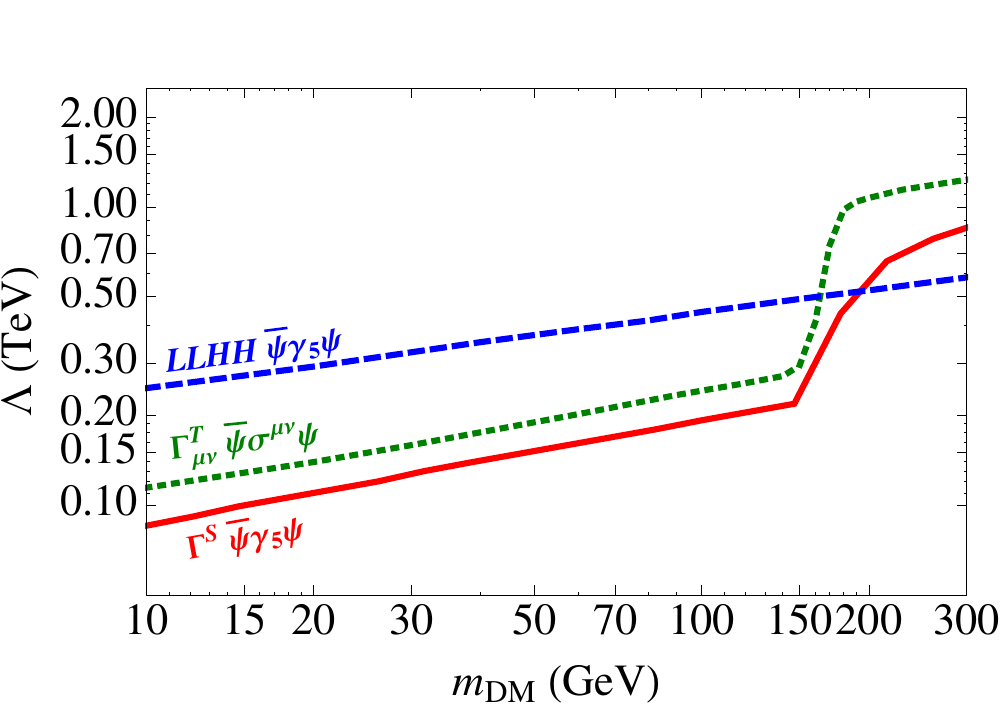}
\caption{The dependence of the EFT cut-off scale $\Lambda$ for scalar (blued dashed line) and tensor (green dotted line) operators \eqref{HFermion} and flavor structure \eqref{cassumpt}  for fermionic DM as a function of DM mass $m_{\rm DM}$ after requiring correct relic density. The solid red line shows $\Lambda$ for leptonic operator \eqref{lnvops}. }
\label{fig:scales}
\end{figure}

Regardless of the detailed flavor structure all these operators do have one feature in common. To obtain correct relic abundance the EFT cut-off scale $\Lambda$ is required to be low, ${\mathcal O}({\rm few~100~GeV})$. The important parameters here are the values of Wilson coefficients $f_\phi, f_\psi^{S,P,T}, f_V$ for bottom quarks in the currents  \eqref{eq:currents} and the value of the Higgs bottom Yukawa coupling (or if this is suppressed, the largest Yukawa coupling among the open annihilation channels). From Higgs data we know that the Higgs bottom Yukawa cannot be significantly larger than the SM value. Using the SM value for $y_b$ we show in Fig.~\ref{fig:scales} the dependence of $\Lambda$ on $m_{\rm DM}$ for scalar and tensor fermionic operators \eqref{HFermion}, setting $f_\psi^S=f_\psi^T=y_b$ as in Eq. \eqref{cassumpt}. Since the annihilation cross section scales as $f_\psi^2/\Lambda^6$  for fermionic DM, taking $f_\psi\sim {\mathcal O}(1)$ still leads to $\Lambda \lesssim 600$ GeV for $m_{\rm DM}<m_h/2$. This means that in any case a viable Higgs portal of light DM using operators in Eq.~\eqref{HFermion} will require new particles with weak scale masses beside DM itself.

Finally, DM can couple to the Higgs through Weinberg-like operator,
\beq
L^iL^jH^kH^l \epsilon_{ik}\epsilon_{jl} \times \mathcal{O}_{\rm dark},
\label{lnvops}
\eeq
where $i,j,k,l$ are $SU(2)_L$ indices, $\epsilon_{ij}$ is the antisymmetric tensor with $\epsilon_{12}=-\epsilon_{21}=1$, and $\mathcal{O}_{\rm dark}$ the DM operator. The lowest dimensional interactions are explicitly,
\begin{subequations}
\begin{align}
\mathcal{H}^0_{\rm eff} &= \frac{g_\phi}{\Lambda^3}L^iL^jH^kH^l \epsilon_{ik}\epsilon_{jl} \times \phi^\dagger \phi, \\
\mathcal{H}^{1/2}_{\rm eff} &= \frac{g_\psi^S}{\Lambda^4}L^iL^jH^kH^l \epsilon_{ik}\epsilon_{jl} \times \overline{\psi}\psi + \frac{g_\psi^P}{\Lambda^4}L^iL^jH^kH^l \epsilon_{ik}\epsilon_{jl} \times i\overline{\psi}\gamma_5\psi, \\
\mathcal{H}^1_{\rm eff} &= \frac{g_V}{\Lambda^3} L^iL^jH^kH^l \epsilon_{ik}\epsilon_{jl} \times V_{\mu}V^{\mu},
\end{align}
\label{lnvops}
\end{subequations}
and similar operators with $\phi^\dagger \phi \to \phi\phi$, $\bar\psi \psi\to \bar \psi^C \psi$ and $\bar\psi \gamma_5 \psi\to \bar \psi^C \gamma_5 \psi$ replacements. 
The operators in Eqs.~(\ref{lnvops}) contribute to neutrino masses at one loop. 
Modulo cancellations, this suppresses all the operators well below the level required for the thermal scattering cross-section to give the observed DM relic density. The only exception is the fermionic DM operator with purely pseudo-scalar interaction ($g_\psi^P$) whose loop contributions to neutrino masses vanish identically by parity invariance, and the $\phi\phi,  \bar \psi^C \psi,  \bar \psi^C \gamma_5 \psi$ type operators if DM carries (conserved) lepton number. 
The resulting invisible Higgs decay governed by the $g_\psi^P$ interaction is very suppressed, that is, ${\cal B}(h \to {\rm DM+DM}+\bar{\nu}\bar{\nu}) \simeq 10^{-7}$  for $m_{\rm DM}=20$ GeV and assuming correct relic DM abundance.
Note that the operator $L^iL^jH^kH^l \epsilon_{ik}\epsilon_{jl} \times i\overline{\psi}\gamma_5\psi$ does induce  DM-nucleon scattering, but only at loop level and the contribution is furthermore proportional to neutrino mass. The DM-nucleon cross section, therefore, is very suppressed. 

The DM annihilation cross section induced by the $L^iL^jH^kH^l \epsilon_{ik}\epsilon_{jl} \times i\overline{\psi}\gamma_5\psi$ operator is given by
\begin{eqnarray}
\sigma_{\psi\bar{\psi} \to \bar{\nu}\bar{\nu}} = \frac{v_{EW}^4(g_\psi^{P})^2}{64\pi \Lambda^8} \frac{s}{\sqrt{1-\beta(m_{\rm DM}^2)}},
\end{eqnarray}
with $\beta(M^2) \equiv 4M^2/s$ and $s\simeq 4 m_{\rm DM}^2$ is the energy in the center of mass frame. The value of $\Lambda$ required to obtain the correct relic density is shown in Fig. \ref{fig:scales} (red solid line), assuming only one neutrino flavor in the final state and setting $g_\psi^p=1$. We observe that the required scale is again low,  i.e. for $m_{\rm DM}=40$ GeV, $\Lambda \simeq 300$ GeV.

In conclusion, our discussion in this section shows that even if the invisible branching ratio of the Higgs is suppressed, viable Higgs portals to light thermal relic DM require new particles with masses of a few $100$~GeV.

\section{Examples of viable Higgs portal models}
\label{sec:4}

 One of the main results of the previous two sections is that Higgs portal models of light DM are still viable, however SM cannot be extended just by DM. Extra light particles are required. The main new ingredient is that the presence of extra light particles increases the DM annihilation cross section, so that correct relic abundance is obtained. Below we show three examples of viable Higgs portal models of light DM. The first two examples illustrate models that match onto EFT discussion of the previous section.  In the first example we  add  to SM and DM an extra electroweak triplet and a singlet (subsection \ref{subsec:A}). This is a realization of a leptophilic model that generates an operator in Eq.~\eqref{lnvops}. The second example is a Two Higgs Doublet Model of type II with an addition of a scalar DM field (subsection \ref{thdm}). It generates EFT operators in Eq.~\eqref{HFermion}. The third example violates EFT assumptions since we add to SM and DM an extra scalar singlet that is lighter than DM (subsection \ref{subsection:C}). As we will see, the value of ${\mathcal B}(h\to{\rm invisible})$ is model dependent. It can be ${\mathcal O}(1)$ as in our example in subsection \ref{subsection:C}, or can be suppressed by the assumed structure of the theory as in the two examples in subsections \ref{subsec:A} and \ref{thdm}.

\subsection{SM + DM with an extra triplet and a singlet}
\label{subsec:A}

In this section, we present a model that could generate the operator $L^iL^jH^kH^l \epsilon_{ik}\epsilon_{jl} \times i\overline{\psi}\gamma_5\psi$. As we will see shortly, it can be done by extending SM particle content by a Dirac fermion DM ($\psi$), an electroweak singlet scalar ($\phi$), and an electroweak triplet scalar ($\Delta$). The extra fields therefore transform under the SM gauge group $SU(3)_C\times SU(2)_L\times U(1)_Y$ as
\beq
\psi\sim (1,1,0), \quad \phi\sim (1,1,0), \quad \Delta\sim (1,3,1).
\eeq
We  use  the notation in which $\Delta$ is represented by the $2\times 2$ matrix,
\begin{eqnarray}
\Delta = \left( \begin{array}{cc}
\Delta^+/\sqrt{2} & \Delta^{++} \\
\Delta^0 & -\Delta^+/\sqrt{2}
\end{array}\right).
\end{eqnarray}

We introduce the following interactions
\beq
\mathcal{L} \supset -\frac{m_\phi^2}{2}\phi^2 - m_\Delta^2 {\rm Tr}\Delta^\dagger \Delta -m_{\rm DM}\bar{\psi}\psi +  \left[ iy\bar{\psi} \gamma_5 \psi \phi + \lambda \phi H^i H^j \epsilon_{ik}\Delta^*_{jk} + f_{ab}L_a^i L_b^j \epsilon_{ik} \Delta_{kj} + {\rm h.c.} \right],
\label{lagr-trip1}
\eeq
where $H$ is the usual SM Higgs doublet, $a,b=1,2,3$ are generation indices, $i,j,k$ are $SU(2)_L$ indices, and $\epsilon_{ij}$ is the antisymmetric tensor. In the above Lagrangian, the $\phi$ is assumed to be a real scalar. Note that we have written only terms relevant to generate the $L^iL^jH^kH^l \epsilon_{ik}\epsilon_{jl} \times i\overline{\psi}\gamma_5\psi$ operator, which is obtained after integrating out $\phi$ and $\Delta$.

It is worth mentioning that one could also consider a variation of the above model in which lepton number is preserved. In this case, the dark matter fermion carries a lepton number -1 and the Lagrangian is modified to
\beq
\mathcal{L} \supset -m_\phi^2\phi^*\phi - m_\Delta^2 {\rm Tr}\Delta^\dagger \Delta -m_{\rm DM}\bar{\psi}\psi +  \left[ y\bar{\psi}^C\psi \phi + \lambda \phi H^i H^j \epsilon_{ik}\Delta^*_{jk} + f_{ab}L_a^i L_b^j \epsilon_{ik} \Delta_{kj} + {\rm h.c.} \right],
\label{lagr-trip2}
\eeq
with $\phi$ complex in this case.

From now on, we shall focus on the model given in Eq. (\ref{lagr-trip1}). The Lagrangian \eqref{lagr-trip1} could be supplemented by several other gauge-invariant terms such as 
\beq
H^T\Delta^\dagger H, \quad \phi{\rm Tr}\Delta^\dagger \Delta, \quad H^\dagger H\phi, \quad H^\dagger H{\rm Tr}\Delta^\dagger\Delta, \quad {\rm Tr}(\Delta^\dagger\Delta)^2, \quad ({\rm Tr}\Delta^\dagger\Delta)^2, \quad H^\dagger \Delta^\dagger\Delta H.
\label{eq:additional}
\eeq
Some of them are already phenomenologically constrained to be small. For instance, $H^T\Delta^\dagger H$ would generate neutrino masses once $\Delta$ is integrated out \cite{Mohapatra:1979ia,Schechter:1980gr,Lazarides:1980nt}. Its coefficient therefore must be very small, much smaller than $m_{\Delta}$. 

By the same reasoning, the term $\mu H^\dagger H\phi$ should be suppressed too.  The simultaneous presences of $f_{ab}L_aL_b\Delta$, $\lambda \phi H^T H\Delta^\dagger$, and $\mu H^\dagger H\phi$ terms breaks lepton number by two units, and as a result the neutrino masses are generated at tree level. To generate unsuppressed Weinberg-like operator \eqref{lnvops} we require $f \sim\lambda \sim 1$ and $m_\phi\sim$ few hundreds GeV, so that $\mu$ needs to be very small, i.e., $\mu \lesssim 1$ eV. Consequently, the $\phi-h$ mixing is extremely suppressed and cannot induce sizeable $h \to {\rm DM+DM}$ decay nor DM-nucleon elastic cross section. The invisible Higgs decay  can thus only occur through the 4-body mode $h\to\bar{\nu}\bar \nu+{\rm DM+DM}$ with branching ratio of $\sim 10^{-6}$ for $m_{DM}=40$ GeV. This number is much too small to be measured in the near future.

\begin{figure}[t]
\centering
\includegraphics[scale=1]{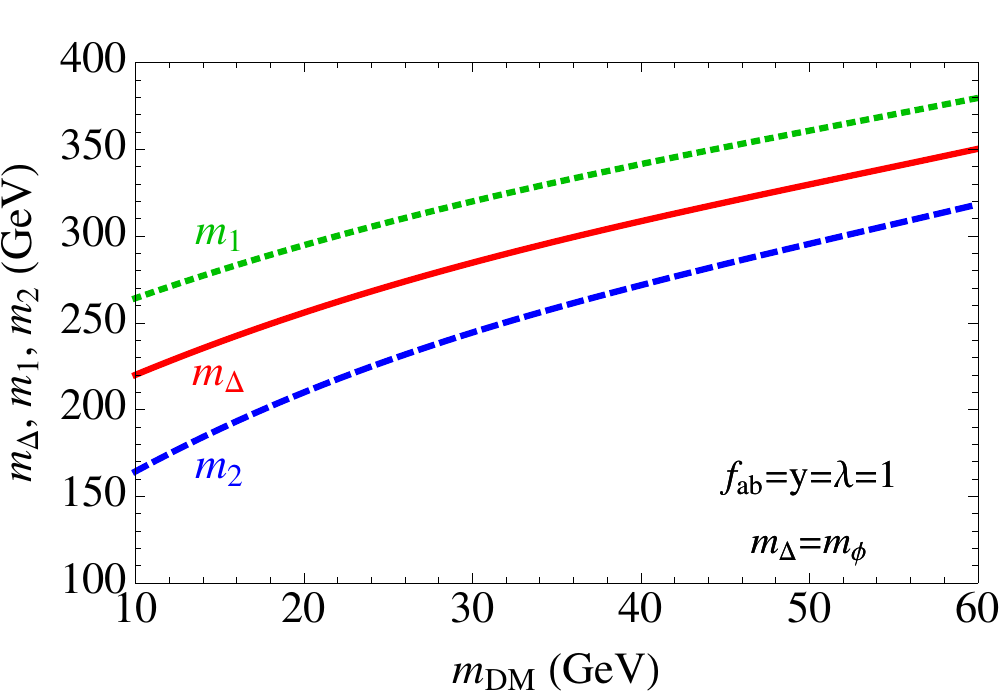}
\caption{The dependence on $m_{\rm DM}$ of the parameter $m_\Delta$ (red solid line) in the SM + DM model with an extra triplet and a singlet Lagrangian \eqref{lagr-trip1}  for which proper relic density is obtained. The masses of physical $\phi-\Delta^0$ mixed states, $m_{1,2}$ are shown as blue dashed and green dotted lines. Other inputs in \eqref{lagr-trip1} are set to  $f_{ab}=y=\lambda=1$ with $m_\phi=m_\Delta$.}
\label{triplet}
\end{figure}

The correct DM relic density is obtained from $\bar{\psi}\psi \to \bar{\nu}\bar\nu$ annihilation that can proceed through $s$-channel $\phi$ and $\Delta^0$ virtual states. The annihilation is unsuppressed as long as there is significant mixing between $\phi$ and $\Delta^0$ states through the $\lambda \phi H^i H^j\epsilon_{ik}\Delta_{jk}^*$ term (after electroweak symmetry breaking). In Fig. \ref{triplet} we show as a function of $m_{\rm DM}$ the  required $m_{\Delta}$ and the masses $m_{1,2}$ of  the two $\phi$--$\Delta^0$ mixed physical states such that the observed DM relic density is generated. The numerical example shown is for maximal mixing, where $m_\phi=m_\Delta$, and we set $f_{ab}=y=\lambda=1$. As anticipated, the required extra states are light, with masses of the order of the weak scale.

The fact that viable Higgs portal models with light DM require additional light states can have phenomenological implications beyond dark matter searches.  In the present model, for instance, there are two charged scalars, $\Delta^{++}$ and $\Delta^+$. These can mediate  lepton flavor violating (LFV) processes such as $\ell_a \to \ell_b\gamma$ and $\ell_a^- \to \ell_b^+ \ell_c^- \ell_d^-$. The radiative decays can arise at one--loop mediated by either $\Delta^+$ or $\Delta^{++}$ particles, with the rate
\begin{eqnarray}
\Gamma(\ell_a \to \ell_b\gamma) = \frac{m_{\ell_a}^5\alpha_{em}}{(24\pi^2)^2} (f^\dagger f)^2_{ab} \left(\frac{1}{8m_{\Delta^+}^2} + \frac{1}{m_{\Delta^{++}}^2} \right)^2,
\end{eqnarray}
where $\alpha_{em}$ is the QED fine-structure constant. The $\ell_a^- \to \ell_b^+ \ell_c^- \ell_d^-$ decay can proceed through tree-level $\Delta^{++}$ exchange, giving
\begin{eqnarray}
\Gamma(\ell_a^- \to \ell_b^+ \ell_c^- \ell_d^-) = \frac{1}{2(1+\delta_{cd})}
\frac{m_{\ell_a}^5}{192\pi^3}\left| \frac{f_{ab}f_{cd}}{m_{\Delta^{++}}^2}\right|^2,
\end{eqnarray}
where $\delta_{cd}$ encodes the symmetry factor for two identical particles in the final state \cite{Nebot:2007bc}.  The resulting bounds on $f_{ab}$ from various LFV processes are given in Table \ref{LFV} for the case of $m_{\Delta^+}=m_{\Delta^{++}}=m_{\Delta}$. (For previous study of LFV in the triplet model, see Refs. \cite{Akeroyd:2009nu,Chun:2003ej,Kakizaki:2003jk}.) For $m_\Delta=220-350$ GeV as required by the relic abundance, the off-diagonal $f_{ab}$ are severely constrained. There are also bounds on diagonal couplings from collider searches. For flavor degenerate case, with $f_{aa}=1$ for $a=1,2,3$, the CMS Collaboration \cite{Chatrchyan:2012ya} reports a bound $m_\Delta > 403$ GeV, which is inconsistent with the relic DM density requirement. The search is less effective for   $f_{\tau\tau}=1$ and $f_{ee}=f_{\mu\mu}=0$,  in which case $\Delta^{--}$ decays exclusively into same-sign tau pairs. The lower limit on $\Delta^{++}$ mass is then $m_\Delta > 204$ GeV  \cite{Chatrchyan:2012ya}, so that correct relic density can still be obtained.

\begin{table}[t]
\centering
\caption{The bounds on LFV couplings $f_{ab}$ of $\Delta$ in Eq. \eqref{lagr-trip1}, following from leptonic LFV decays. The experimental $95\%$ C.L. upper bounds are from \cite{pdg}, except for $\mu \to e\gamma$ which is from \cite{Adam:2013mnn}. We set $m_{\Delta^+}=m_{\Delta^{++}}=m_{\Delta}$.}
\label{LFV}
\begin{tabular}{lcc}
\hline\hline
Process \hspace{1cm}&  Branching ratio bound \hspace{1cm}& \hspace{0cm} Bounds on $f_{ab}$\\ \hline\hline
$\mu^- \to e^+e^-e^-$  & $1.0 \times 10^{-12}$ & $|f_{ee}f_{e\mu}|<2.8 \times 10^{-5}~\left({m_\Delta}/{\rm TeV} \right)^2$ \\
$\tau^- \to e^+e^-e^-$  & $2.7 \times 10^{-8}$ & $|f_{ee}f_{e\tau}|< 0.01 ~\left({m_\Delta}/{\rm TeV} \right)^2$ \\
$\tau^- \to e^+e^-\mu^-$  & $1.8 \times 10^{-8}$ & $|f_{e\mu}f_{e\tau}|< 0.007~\left({m_\Delta}/{\rm TeV} \right)^2$ \\
$\tau^- \to e^+\mu^-\mu^-$  & $1.7 \times 10^{-8}$ & $|f_{\mu\mu}f_{e\tau}|< 0.009~\left({m_\Delta}/{\rm TeV} \right)^2$ \\
$\tau^- \to \mu^+e^-e^-$  & $1.5 \times 10^{-8}$ & $|f_{ee}f_{\mu\tau}|< 0.008~\left({m_\Delta}/{\rm TeV} \right)^2$ \\
$\tau^- \to \mu^+\mu^-e^-$  & $2.7 \times 10^{-8}$ & $|f_{e\mu}f_{\mu\tau}|<0.009~\left({m_\Delta}/{\rm TeV} \right)^2$ \\
$\tau^- \to \mu^+\mu^-\mu^-$  & $2.1 \times 10^{-8}$ & $|f_{\mu\mu}f_{\mu\tau}|<0.01 ~\left({m_\Delta}/{\rm TeV} \right)^2$ \\
$\mu \to e\gamma$  & $5.7 \times 10^{-13}$ & $|f_{\mu a}^*f_{ae}| < 2.7 \times 10^{-4} ~\left({m_\Delta}/{\rm TeV} \right)^2$ \\
$\tau \to e\gamma$  & $3.3 \times 10^{-8}$ & $|f_{\tau a}^*f_{ae}| < 0.15 ~\left({m_\Delta}/{\rm TeV} \right)^2$ \\
$\tau \to \mu\gamma$  & $4.4 \times 10^{-8}$ & $|f_{\tau a}^*f_{a\mu}| < 0.18 ~\left({m_\Delta}/{\rm TeV} \right)^2$ \\
\hline\hline
\end{tabular}
\end{table}

\subsection{2HDM-II + DM}
\label{thdm}

Our next example of a viable Higgs portal DM is a type II Two-Higgs-Doublet-Model (2HDM-II) supplemented by an extra singlet scalar -- the DM. This is the simplest realization of the fermionic operators in Eq. \eqref{HFermion},  discussed in the previous Section assuming EFT. While phenomenologically viable, the model does have two ad-hoc features. The invisible Higgs decay width is suppressed by dialling down the appropriate dimensionless parameter, while direct DM detection bounds are avoided by fine-tuning the parameters so that two competing operator contributions cancel to a large extent. 

The detailed structure of the model is as follows. The particle content consists of SM fermions, two Higgs doublets, $H_1$ and $H_2$, and an extra real scalar $S$. Under SM gauge group, these scalars transform as
\begin{eqnarray}
H_1\sim \left(1,2,1/2\right)\,, \quad H_2\sim \left(1,2,1/2\right)\,, \quad S\sim \left(1,1,0\right)\,.
\end{eqnarray}
The singlet $S$ is assumed to be $Z_2$ odd and is identified as DM. The Yukawa interactions of the two doublets are assumed to be the same as in type II 2HDM; $H_1$ couples to $d_R$ and $e_R$, while $H_2$ only couples to $u_R$,
\begin{eqnarray}
\mathcal{L}_Y = -Y_u \overline{Q} \tilde{H}_2 u_R - Y_d \overline{Q} H_1 d_R - Y_\ell \overline{L} H_1 e_R + {\rm h.c.},
\end{eqnarray} 
where $\tilde{H}_i\equiv i\sigma_2H_i^*$ and $H_i = \left(H_i^+,(v_i+h_i+i\chi_i)/\sqrt{2}\right)$. DM couples directly to the two Higgs doublets,
\begin{eqnarray}
\mathcal{L} &\supset& \frac{\lambda_{S1}}{2} S^2 (H_1^\dagger H_1 ) + \frac{\lambda_{S2}}{2} S^2(H_2^\dagger H_2).
\end{eqnarray}
For suitable choices of parameters, these interactions allow for large enough DM annihilation cross section and as a result can accommodate the observed relic abundance. 

After electroweak symmetry breaking three out of eight real degrees of freedom in $H_1$ and $H_2$ are absorbed as longitudinal components of $W^{\pm}$ and $Z$ bosons (for reviews see e.g.~\cite{Gunion:2002zf, Branco:2011iw}). The remaining 5 degrees of freedom consist of two CP-even scalars $h$ and $H$, 
\begin{eqnarray}
\left( \begin{array}{c}
H \\ h \end{array} \right)
&=& \left( \begin{array}{cc}
\cos\alpha & \sin\alpha \\ -\sin\alpha & \cos\alpha \end{array} \right)
\left( \begin{array}{c}
h_1 \\ h_2 \end{array} \right), 
\end{eqnarray}
a CP-odd scalar $A\equiv -\chi_1\sin\beta+\chi_2\cos\beta$, and a pair of charged scalars $H^{\pm}\equiv -H_1^\pm \sin\beta+H_2^\pm\cos\beta$. Here $\tan\beta \equiv v_2/v_1$ is the ratio of $H_{2,1}$ condensates with $v_{EW}\equiv \sqrt{v_1^2+v_2^2}$. It is $h$ that we identify as the newly discovered particle with 125 GeV mass. The interactions of the CP-even scalars, $h,H$, with the SM fermions and gauge bosons are given by
\begin{eqnarray}
\mathcal{L} &\supset& - \sum_{f=u,d,\ell} \left(\frac{r_f m_f}{v_{EW}}h  + \frac{R_f m_f}{v_{EW}}H\right) \overline{f}f
+ g\sin(\beta-\alpha)\left( m_W W_\mu^+W^{\mu -} + \frac{m_Z}{2c_W} Z_\mu Z^\mu \right)h \nonumber \\
&&+ g\cos(\beta-\alpha)\left( m_W W_\mu^+W^{\mu -} + \frac{m_Z}{2c_W} Z_\mu Z^\mu \right)H,
\label{cp-even-sm}
\end{eqnarray}
with
$r_u = \cos\alpha\csc\beta, 
r_d=r_\ell=-\sin\alpha\sec\beta,$
$R_u = \sin\alpha\csc\beta, 
R_d=R_\ell=\cos\alpha\sec\beta$. 
After electroweak symmetry breaking there are also trilinear couplings of $h,H$ with the DM,
\begin{eqnarray}
\mathcal{L} \supset \frac{g_{SSh}}{2}v_{EW} hS^2 + \frac{g_{SSH}}{2} v_{EW} HS^2, 
\end{eqnarray}
where
\begin{eqnarray}
g_{SSh} &=& \lambda_{S1}\sin\alpha \cos\beta -\lambda_{S2} \cos\alpha \sin\beta, \nonumber \\
g_{SSH} &=& -\lambda_{S1} \cos\alpha \cos\beta - \lambda_{S2} \sin\alpha \sin\beta.
\end{eqnarray}

\begin{figure}[t]
\centering
\includegraphics[scale=1.4]{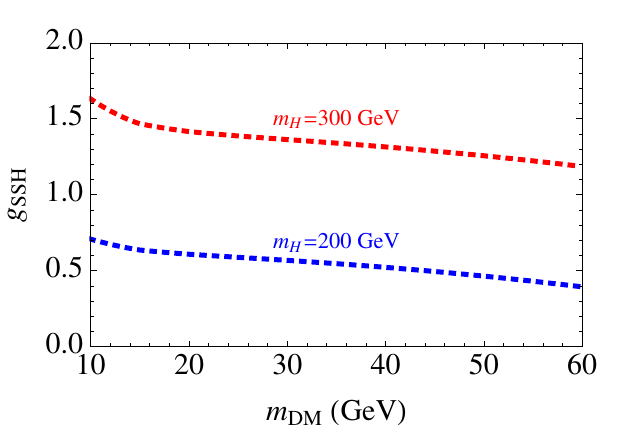}
\caption{The value of $g_{SSH}$ that gives the observed DM relic density in 2HDM-II models with extra singlet, as function of DM mass, $m_{\rm DM}$, for the case where the invisible decay width of the Higgs and the DM-proton scattering cross section both vanish. Two choices of the heavy CP-even Higgs mass, $m_H=200, 300$ GeV are shown.}
\label{rd-thdm}
\end{figure}

DM annihilation into a pair of SM fermions, $SS\to \bar f  f$,  is mediated by both CP-even scalars, $h$ and $H$ and is proportional to $\sigma_{\rm ann}\propto (g_{SSh}/m_h^2+g_{SSH}/m_H^2)^2$. For light DM the $g_{SSh}$ coupling also leads to ${\mathcal B}(h\to SS)$. As we show below the bounds on invisible decay width of the Higgs require $g_{SSh}<0.01$. Correct relic abundance then requires $g_{SSH}\sim {\mathcal O}(1)$, see Fig. \ref{rd-thdm}.

Similarly, DM--nucleon scattering cross section also receives contributions from both $h$ and $H$ exchanges, 
\begin{eqnarray}
\sigma_p^{SI} &=& \frac{m_p^4}{4\pi(m_{\rm DM}+m_p)^2m_H^4}\left(\sum_q c_q f_q^p\right)^2,
\label{dd-2hdm}
\end{eqnarray}
where
\begin{eqnarray}
c_{u,c,t} &=& g_{SSh} (m_H/m_h)^{2} \cos\alpha\csc\beta  + g_{SSH} \sin\alpha\csc\beta, \nonumber \\
c_{d,s,b} &=& -g_{SSh} (m_H/m_h)^{2} \sin\alpha\sec\beta + g_{SSH} \cos\alpha\sec\beta\,,
\end{eqnarray}
while the relevant nuclear form factors $f_q^p$ are listed in Eq.~\eqref{nucl-form-fact}\,.
The $h$ and $H$ contributions may interfere destructively. In fact, $\sigma_p^{SI}$ vanishes completely, if
\begin{eqnarray}
\frac{g_{SSh}}{g_{SSH}} = \frac{m_h^2}{m_H^2} \frac{(f^p_u+f^p_c+f^p_t)\sin\alpha\cos\beta + (f^p_d+f^p_s+f^p_b) \cos\alpha\sin\beta}{(-f^p_u-f^p_c-f^p_t)\cos\alpha\cos\beta+(f^p_d+f^p_s+f^p_b) \sin\alpha\sin\beta}\,.
\label{dd-cond}
\end{eqnarray}
Note that it is possible to fulfill this requirement even if $g_{SSh}=0$. Then $\mathcal{B}(h \to SS)=0$, while Eq. \eqref{dd-cond} gives
\begin{eqnarray}
\frac{\tan\alpha}{\tan\beta} = -\frac{f^p_d+f^p_s+f^p_b}{f^p_u+f^p_c+f^p_t}\,.
\label{inv-cond}
\end{eqnarray}
As we will show below the pseudo-decoupling limit, $\beta-\alpha=\pi/2$, where the couplings of the Higgs to $W$ and $Z$ are the SM ones, c.f. Eq.~(\ref{cp-even-sm}), is preferred by recent Higgs data. In this limit Eq.~\eqref{inv-cond} then completely fixes the value of $\tan\beta$; i.e., using the values of nuclear form factors in  Eq.~\eqref{nucl-form-fact} one obtains $\tan\beta\simeq0.61$. 

In the limit  where ${\cal B}(h \to SS)$ vanishes, the relic abundance is set by DM annihilation with the heavy CP-even Higgs boson $H$  in the {\it s}-channel. In Fig.~\ref{rd-thdm}, we plot the coupling $g_{SSH}$ giving the correct relic abundance as a function of DM mass, $m_{\rm DM}$, for two sample values of heavy CP-even Higgs boson masses, $m_{H}=200,300$~GeV. We also set $\tan\beta=0.61$ such that $\sigma_p^{SI}$ vanishes. For heavier $H$ a larger value of $g_{SSH}$ coupling is needed. Perturbativity  therefore bounds $m_H$ from above, with $m_H\lesssim 850$ GeV for $g_{SHH}\lesssim 4\pi$ (and $m_H\lesssim 450$ GeV for $g_{SHH}\lesssim 4$). Note that in this case $H$ decays invisibly practically 100\% of the time. In principle $H$ can be directly searched in the process of associated production with a $Z$ boson (see, e.g., a recent ATLAS analysis of $pp\to Zh\to l^{+}l^{-}\textrm{invisible}$ \cite{ATLAS-inv}). The challenge is that in the limit $\beta-\alpha=\pi/2$, the  couplings of $H$ to gauge bosons vanish. As a result, the heavy Higgs boson in this scenario can easily escape such collider searches. On the other hand, $H$ also couples to SM fermions with roughly SM strengths, thus making $gg\to H(t\bar t)$ the dominant production mechanisms at the LHC. Especially in the second case, the dominant decay mode $H\to $DM+DM then leads to the interesting $t\bar t + E_T^{\rm miss}$ signature. For $m_H=200,~300$\,GeV, we find using~\cite{Heinemeyer:2013tqa} the cross-section estimates of $\sigma_{t\bar t E_{T}^{\rm miss}} = 29\,{\rm fb},~7.7$\,fb at 8~TeV and $\sigma_{t\bar t E_{T}^{\rm miss}} = 150\,{\rm fb},~51$\,fb at 14~TeV LHC, respectively. Given these small cross-sections, also compared to irreducible SM ($t\bar t+Z$) backgrounds~\cite{Lazopoulos:2008de}, the search remains challenging for the foreseeable future.
On the other hand, interesting mono-jet plus missing transverse energy signature would come from $gg \to H+$jet. Using this particular signature, a dedicated analysis for the SM Higgs boson invisible decay was performed in~\cite{Englert:2011us}. The upper limit on $\mu_{Hj}\equiv \sigma_{gg\to Hj} \times {\mathcal B}(H\to\textrm{inv})/\sigma_{gg\to Hj}^{SM}$ at  $95\%$~C.L.  was found to be $\mu_{Hj} < 25\, (50)$  for $200$~GeV ($300$~GeV) Higgs boson using just $1$~fb$^{-1}$ of data at 7 TeV. It might be possible for $14$~TeV LHC to probe the prediction of this model, $\mu_{Hj}=R_u^2=2.7$.

\begin{figure}[t]
\centering
\includegraphics[scale=0.9]{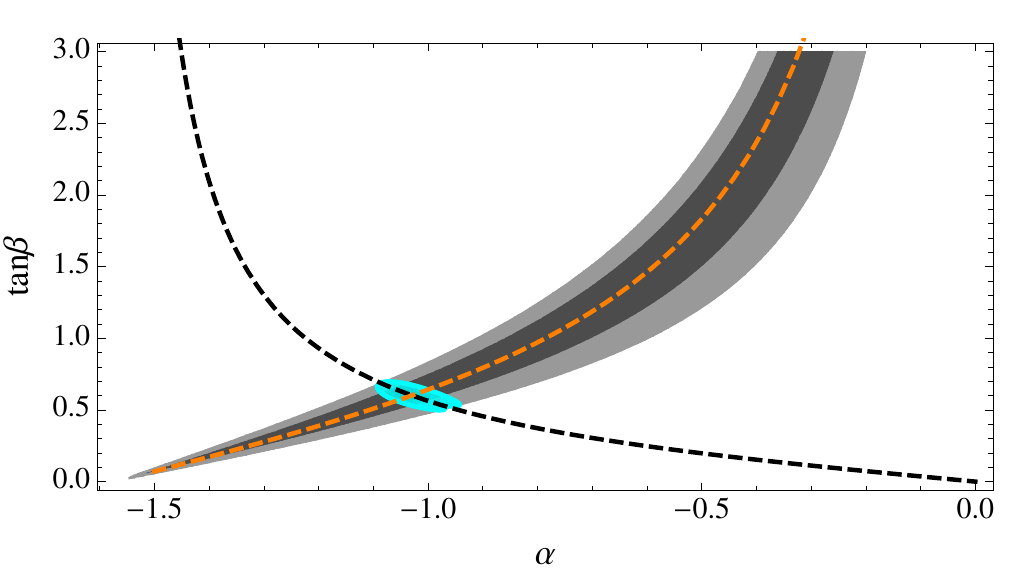}
\caption{The $68.3\%$ and $95.5\%$ C.L.   parameter regions in 2HDM-II with an extra singlet that are allowed by the Higgs signal strength data are shown in dark grey and light grey, respectively. Orange-dashed curve correspond to $\beta-\alpha=\pi/2$. Black-dashed curve correspond to Eq. (\ref{inv-cond}). The $95.5\%$ C.L. region allowed by the Higgs data together with direct DM detection bound from XENON100 is shown in cyan. For definiteness we assume $m_H=200\,$GeV, $m_S=40$ GeV and $g_{SSH}$ such that the proper DM thermal relic is obtained.}
\label{fit-thdm}
\end{figure}

Finally, we assess the quantitative impact of existing Higgs measurements on the model's parameter space by performing a fit to the latest LHC Higgs data assuming that $h$ is the newly discovered Higgs resonance (for details see Appendix \ref{app:Higgs:fit}). 
The partial decay widths normalized to the SM ones are given by
\beq
\frac{\Gamma_{h\to WW,ZZ}}{\Gamma_{h\to WW,ZZ}^{SM}}=\sin^2(\beta-\alpha) \equiv r_{V}^{2}\,, \quad \frac{\Gamma_{h\to bb}}{\Gamma_{h\to bb}^{SM}}=r_{d}^{2}\,, \quad \frac{\Gamma_{h\to\tau\tau}}{\Gamma_{h\to\tau\tau}^{SM}}=r_{l}^{2}\,, \quad \frac{\Gamma_{h\to\gamma\gamma}}{\Gamma_{h\to\gamma\gamma}^{SM}}=\left|-1.28r_{V}+0.283r_{u}\right|^{2}\,,
\eeq
while the normalized production rates are 
\begin{equation}
\frac{\sigma_{ggF}}{\sigma_{ggF}^{SM}}=\left|1.06r_{u}+(-0.06+\imath0.09)r_{d}\right|^{2},\;\;\;\;\frac{\sigma_{VBF+VH}}{\sigma_{VBF+VH}^{SM}}=r_{V}^{2}\,.
\end{equation}
In the Higgs signal strengths, $\mu_i$, one measures the product of cross section and Higgs branching ratios. Therefore in all the signal strengths the total Higgs decay width enters. This can be modified by the invisible decay width of the Higgs, and as a result one is quite sensitive to it. Normalized to the SM the total width is given by
\beq
\hat{\Gamma}\equiv\frac{\Gamma_{total}}{\Gamma_{total}^{SM}}=\frac{0.569r_{d}^{2}+0.252r_{V}^{2}+0.063r_{l}^{2}+0.085\frac{\sigma_{ggF}}{\sigma_{ggF}^{SM}}+0.026r_{u}^{2}}{1-{\cal B}(h\to S S)}\,.
\eeq
Numerical values for loop functions in $h\to\gamma\gamma$ and $h\to g g$ are taken from \cite{Djouadi:2005gi}, while SM branching ratios for $m_h=125\,$GeV Higgs boson are taken from \cite{Dittmaier:2012vm}. 
In our model all the Higgs 
signal strengths $\mu_{i}$ depend on three parameters, $\alpha$, $\beta$ and 
${\cal B}(h\to S S)$.
Fig.~\ref{fit-thdm} shows the $68.3\%$ and $95.5\%$ C.L. allowed region in the parameter space ($\alpha$, $\tan\beta$) obtained from a global fit after marginalizing over 
${\cal B}(h\to S S)$. The allowed parameter space is constrained to a very narrow region around $\beta-\alpha=\pi/2$. We also derive the bound on invisible branching ratio of the Higgs by marginalizing over $\alpha$ and $\tan \beta$. We get 
${\cal B}(h\to S S)<0.3$ at $95.5\%$ C.L., which implies that $g_{SSh}<0.01$ for DM mass up to $m_h/2$. We emphasize that ${\cal B}(h\to {\rm invisible})$ is a free parameter in this model, and can be both close to present experimental bound or much smaller, depending on the derived dimensionless parameter $g_{SSh}$.

Finally, we combine the Higgs data and $90\%$ C.L. upper bound on spin-independent DM-nucleon cross section from XENON100 \cite{aprile:2012} into a single $\chi^2$. For illustration we fix $m_H=200\,$GeV, $m_S=40\,$GeV and $g_{SSH}$ to value determined by relic density. The DM scattering cross section $\sigma_{p}^{SI}$ and the signal strength rates $\mu_i$ are expressed in terms of three fitting parameters $\alpha$, $\beta$ and $g_{SSh}$. After marginalizing over $g_{SSh}$, we obtain the $95.5\%$ C.L. allowed region in $(\alpha,\tan\beta)$ plane, shown as cyan region in Fig.~\ref{fit-thdm}. Marginalizing analogously over $\alpha$ and $g_{SSh}$, we find $\tan\beta=(0.61\pm0.03)$.

\subsection{SM + DM with extra scalar singlet}
\label{subsection:C}

In our final example of a viable Higgs portal model of DM
we add to the SM two real scalars, $\phi$ and $S$ (for existing studies of similar models see~\cite{Abada:2011qb}). Under the SM gauge group both scalars therefore transform as
\beq
\phi\sim (1,1,0)\,, \quad S\sim (1,1,0)\,.
\eeq
The singlet $S$ is the DM candidate, odd under $Z_2$, 
while 
$\phi$ is even.  The resulting scalar potential
is 
\begin{eqnarray}
V &=& m_H^2 H^\dagger H + \frac{m_2^2}{2}\phi^2 + \frac{m_3^2}{2}S^2 + \kappa m_2^3 \phi+\frac{\lambda_1}{2}(H^\dagger H)^2 + \frac{\lambda_2}{8}\phi^4 + \frac{\lambda_3}{8}S^2  \nonumber \\
&& + \frac{\lambda_4}{2} H^\dagger H \phi^2 + \frac{\lambda_5}{2} H^\dagger H S^2 + \frac{\lambda_6}{4} \phi^2 S^2  + \frac{\mu_1}{2}\phi^3 + \mu_2 H^\dagger H \phi + \frac{\mu_3}{2} S^2\phi\,, \label{eq:Vscalar}
\end{eqnarray}
while the Yukawa interactions take the usual form
\begin{eqnarray}
-\mathcal{L}_Y = Y_u \overline{Q} \tilde{H} u_R + Y_d \overline{Q} H d_R + Y_\ell \overline{L} H e_R + {\rm h.c.}\,.
\end{eqnarray}
For simplicity, we assume that  $\phi$ does not acquire a vacuum expectation value by appropriately adjusting the parameter $\kappa$ (this has no relevant phenomenological consequences apart from simplifying our discussion).
The scalar mass matrix is given by
\begin{eqnarray}
M_{\textrm{sc}}^{2}=\left(\begin{array}{cc}
m_{h}^{2} & \mu_2 v_{EW}\\
\mu_2 v_{EW} & m_{\phi}^{2}
\end{array}\right),
\end{eqnarray} 
where $m_{h}^{2}= \lambda_1 v_{EW}^{2}$ and $m_{\phi}^{2}=m_{2}^{2}+\lambda_{4} v_{EW}^{2}/2$. Parameter $\mu_2$ induces mixing between $h$ and $\phi$, so that  the physical neutral scalars $h_1,h_2$ are given by
\begin{eqnarray}
h_1 &=& h\cos\alpha  + \phi \sin\alpha\,,\nonumber \\
h_2 &=& -h\sin\alpha  + \phi \cos\alpha\,,
\end{eqnarray}
with the mixing angle given by
\begin{eqnarray}
\tan2\alpha=\frac{2\mu_2 v_{EW}}{m_{h}^{2}-m_{\phi}^{2}}\,.
\end{eqnarray}
We will assume that $m_{h_1}/2>m_{S}>m_{h_2}$ with $m_{h_1}=125\,$GeV.

The couplings of $h_1$ ($h_2$) to the SM fields are the same as for the SM Higgs boson except that they are rescaled by $\cos\alpha$ ($\sin\alpha$). The mixing angle $\alpha$ has been constrained by LEP  \cite{Barate:2003sz}, so that  at 95\% C.L. $\left|\sin\alpha\right|<0.13$ for $m_{h_2}=20\,$GeV and $\left|\sin\alpha\right|<0.2$ for $m_{h_2}=50\,$GeV. On the other hand, $\sin\alpha$ also has to be greater than $~10^{-8}$, otherwise $h_2$ is sufficiently long lived that it escapes the detector.
 For $\sin\alpha\thicksim10^{-4}$ the $h_2$ particle travels less than a few $\mu$m before decaying and can be searched for using displaced vertices. Note that the branching ratios of $h_2$ are not affected by $\sin\alpha$ and are the same as they would be for the SM Higgs with $m_{h_2}$ mass. For instance, for $m_{h_2}=20$ GeV the dominant branching ratio is ${\cal B}(h_2 \to b\bar{b})\sim85\%$.

\begin{figure}[t]
\centering
\includegraphics[scale=1.45]{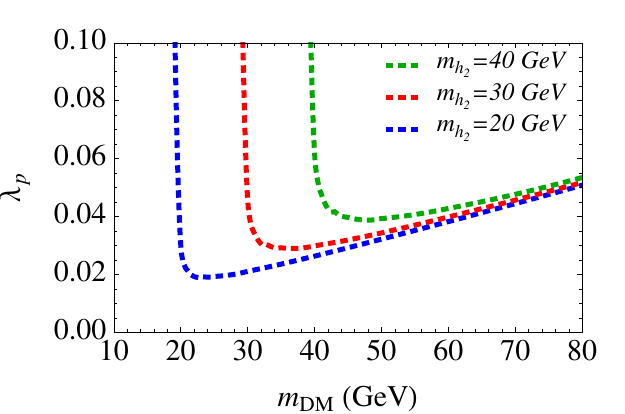}
\caption{Coupling $\lambda_p$ for which the proper relic abundance is obtained in the model with an extra scalar singlet \eqref{eq:Vscalar}.}
\label{extrasin}
\end{figure}

The relic abundance is set by the dominant DM annihilation process $SS\to h_2 h_2$, with the annihilation cross section given by
\beq
\sigma_{SS\to h_{2}h_{2}}=\frac{\lambda^{2}_{p}}{32\pi s}\frac{\sqrt{1-4m_{h_{2}}^{2}/s}}{\sqrt{1-4m_{S}^{2}/s}}\,,
\eeq
where $\lambda_p=\lambda_{6}\cos^{2}\alpha+\lambda_{5}\sin^{2}\alpha$. The values of $\lambda_p$ for which the correct relic abundance is obtained are shown in Fig.~\ref{extrasin} as a function of DM mass, $m_{\rm DM}$, for three choices of light scalar mass $m_{h_2}$. 
Note that $\lambda_p$ that governs the relic abundance is different from $\lambda_h=\lambda_5\cos\alpha-\lambda_6\sin\alpha$ that governs the invisible Higgs branching ratio, ${\mathcal B}(h\to {\rm invisible})$. The relic abundance and invisible decay width of the Higgs are thus decoupled in this Higgs portal model. 

\begin{figure}[t]
\centering
\includegraphics[scale=1.5]{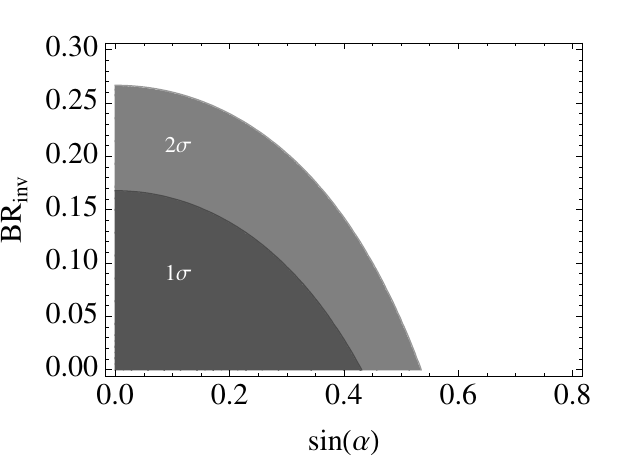}
\caption{Constraints from Higgs signal strengths of the Higgs portal model of light DM with an extra singlet. $1\sigma$ and $2\sigma$ constraints on $\sin\alpha$ and ${\cal B}(h\to {\rm invisible})$ are show as dark and lighter grey regions, respectively.}
\label{fit-extrasin}
\end{figure}

Next, we perform fit to the latest available LHC Higgs data. Unlike the 2HDM-II case, Section \ref{thdm}, here all the Higgs ($h_1$) signal strengths are rescaled by common factor $\cos^2\alpha$. Also, there are additional contributions to the total Higgs decay width coming from $h_1\to h_2h_2$ and $h_1\to S S$. 
The Higgs signal strengths, therefore, are given by
\beq
\mu^{h\to SM}=\cos^{2}\alpha(1-\Delta\hat{\mathcal B})\,,
\eeq
with $\Delta\hat{\mathcal B}\equiv{\cal B}(h_1 \to SS)+{\cal B}(h_1\to h_2 h_2)$. A direct bound on invisible Higgs decay width from ATLAS analysis of $pp\to Zh\to l^{+}l^{-}\textrm{invisible}$ \cite{ATLAS-inv}, is added to $\chi^2$ as
\beq
\chi_{inv}^{2}=\left(\frac{\cos^{2}\alpha {\cal B}(h_1 \to SS)+0.19}{0.43}\right)^{2},
\eeq
assuming that $\sin\alpha \gtrsim 10^{-4}$ so that $h_2$ decays instantaneously.
We then take $\sin\alpha$, ${\cal B}(h_1 \to SS)$ and ${\cal B}(h_1\to h_2 h_2)$ as fitting parameters. We obtain $95.5\%$ C.L. bounds on each parameter to be $\left|\sin\alpha\right|<0.5$, ${\cal B}(h_1\to h_2 h_2)<0.24$ and ${\cal B}(h_1 \to SS)<0.22$. Note that the bound on $\sin\alpha$ obtained from this fit is less stringent than the LEP limit. In Fig.~\ref{fit-extrasin}, we show $68.3\%$ and $95.5\%$ C.L. allowed region in the parameter space of $\sin\alpha$ and ${\cal B}(h_1 \to SS)$, after marginalizing over ${\cal B}(h_1\to h_2 h_2)$. If $\sin\alpha$ is very small, so that $h_2$ escapes the detector, then we obtain ${\cal B}(h_1 \to {\rm invisible})<0.22$.

Since there is an extra light scalar state, there are interesting collider signatures beside the invisible decay width of the Higgs. The Higgs can also decay to two light scalars, $h_1\to h_2 h_2$, where $h_2$ decays to $b\bar b$ pairs. These  decay chains can then be searched for using associated $hZ$ or $hW$ production with four $b$-tagged jets in the final state (possibly originating from two displaced secondary vertices, see also~\cite{Halyo:2013yfa}) combining to the Higgs mass. As discussed above, the $h_1\to h_2 h_2$ branching ratio can be sizeable, of ${\mathcal O}(20\%)$.

\section{Conclusions}
\label{sec:conclusions}

In this paper we have extended the analysis of Higgs portal models of DM by including higher dimensional operators. We focused on the case where DM is light, so that $h\to$ DM+DM decays are kinematically allowed. The main difference between the minimal Higgs portals and the case where higher dimensional operators dominate, is that there is now a new scale $\Lambda$ in the problem. In fact, already for minimal Higgs portal with fermionic DM one is forced to introduce a dimensionful scale $\Lambda$ since the Higgs couplings then require at least dimension 5 operators. We arrive at the following general conclusions

\begin{itemize}
\item
 First assume that an EFT description of SM+DM as the only weak scale dynamical degrees of freedom is valid and all dimensionless coefficients are ${\mathcal O}(1)$. If $h\to {\rm DM+DM}$ is discovered close to its present experimental limit, at the order of ${\mathcal O}({\rm few~10\%})$, then DM cannot be a thermal relic, or its relic density must be controlled by interactions not involving the Higgs field.
 \item
 Higgs portal to DM is still possible if either EFT is not valid or if  ${\mathcal B}(h\to {\rm invisible})$ is  suppressed below naive counting estimate (or both). In both cases there need to be other light particles, with masses below ${\mathcal O}({\rm few~100~GeV})$.
 \end{itemize}

We demonstrate this with three examples of viable Higgs portal models of light DM, (i) the SM extended by DM scalar along with electroweak triplet and singlet  (subsection \ref{subsec:A}), (ii) a Two Higgs Doublet Model of type II with an addition of scalar DM field (subsection \ref{thdm}), (iii) SM with DM and an extra scalar singlet that is lighter than DM (subsection \ref{subsection:C}). All the examples share the feature that the invisible Higgs branching ratio and the relic abundance are decoupled and are governed by different parameters. Furthermore, only in example (ii) the dominant DM annihilation channel is to $b\bar b$ pairs as in the simplest Higgs portal models. As a result this model also requires tuned cancellation to avoid direct DM detection constraints. 

Since the Higgs portals of DM require additional light particles, there may be interesting phenomenological consequences.  Indeed, non-trivial dynamics 
taking place below the TeV could leave significant footprints in low energy precision observables, or could be directly detected at high-energy collider 
experiments. For instance, the charged scalars in example (i) can lead to lepton flavor violating decays, in example (ii) the heavy Higgs decay is dominated by the invisible channels, while in (iii) the Higgs decays to four $b$ jets at the level of a few tens of percent are possible. 

\begin{acknowledgments}
A.G., J.J. and J.F.K were supported in part by the Slovenian Research Agency. J.Z. was supported in part by the U.S. National Science Foundation under CAREER Grant PHY-1151392.
\end{acknowledgments}

\appendix

\section{Relic density and direct detection}
\label{sec:3}

The DM relic abundance is found by solving the following Boltzmann equation,
\beq
\frac{dY}{dx}= \frac{1}{3H}\frac{ds}{dx} \left< \sigma v \right>(Y^2-Y_{\rm eq}^2)\,,
\eeq
where $H$ is the Hubble constant, $x\equiv m_{\rm DM}/T$ with $m_{\rm DM}$ the DM mass, and
$Y\equiv n/s$ with $n$ and $s$ the  number density and entropy density respectively. The thermal average of the annihilation cross section  
is  given by \cite{gondolo91}
\begin{eqnarray}
\left< \sigma v \right> = \int_{\epsilon_{th}}^\infty d\epsilon \frac{2x}{K_2(x)^2}\sqrt{\epsilon}(1+2\epsilon)K_1(2x\sqrt{1+\epsilon})\sigma v\,.
\end{eqnarray}
Here $K_i(x)$ is the $i-$th order modified Bessel function of the second kind. The parameter $\epsilon$ is the kinetic energy per unit mass defined as $\epsilon \equiv (s-4m_{\rm DM}^2)/(4m_{\rm DM}^2)$, while $\epsilon_{th}$ is the threshold kinetic energy per unit mass. It is $\epsilon_{th}=0$ if $2m_{\rm DM} \geq m_3+m_4$, and $\epsilon_{th}=(m_3+m_4)^2/(4m_{\rm DM}^2)-1$ if $2m_{\rm DM}<m_3+m_4$, with $m_3$ and $m_4$ the masses of the final state particles.

In the early universe, DM is assumed to be in equilibrium. Once the temperature drops below the DM mass, $Y_{\rm eq}$ is exponentially suppressed. When the freeze-out temperature is reached, the equilibrium is no longer maintained.
As the result, one can integrate the Boltzmann equation to determine relic abundance \cite{gondolo91,griest-seckel91}
\beq
\Omega h^2 = \frac{1.07 \times 10^9~{\rm GeV}^{-1}}{\sqrt{g_*}M_{\rm Pl}} \left(\int_{x_f}^\infty dx \frac{\left< \sigma v \right>}{x^2} \right)^{-1}\,,
\eeq
where $M_{\rm Pl}=1.22 \times 10^{19}$ GeV is the Planck mass, and $g_*$ is the number of effective relativistic degrees of freedom at freeze-out.
The freeze-out temperature $T_f$ is determined through ($x_f\equiv m_{\rm DM}/T_f$)
\beq
x_f = \ln \frac{0.038gM_{\rm Pl}m_{\rm DM} \left<\sigma v\right>}{\sqrt{g_* x_f}}\,,
\eeq
with $g$ the number of DM degrees of freedom.

We review next the calculation of direct DM detection bounds. The operators given in Eqs. (\ref{GaugeOp}) and (\ref{HFermion}) lead to the DM-quark interactions which then induce the scattering of DM on nuclei.
For operators in Eq.~(\ref{GaugeOp}), the DM-nucleon cross sections are found to be ($N=p,n$)
\begin{eqnarray}
\sigma^{\phi SI}_{p,n} &=&  \frac{8G_F^2}{\pi}c^2_\phi\left(\frac{v_{EW}}{\Lambda}\right)^4 \mu^2_{\phi N}(2 Y_{u,d}+Y_{d,u})^2,  \nonumber \\
\sigma^{\psi SI}_{p,n} &=& \frac{G_F^2}{2\pi}(c_\psi^L+c_\psi^R)^2\left(\frac{v_{EW}}{\Lambda}\right)^4\mu^2_{\psi N}(2 Y_{u,d}+Y_{d,u})^2,   \nonumber \\
\sigma^{\psi SD}_{p,n} &=& \frac{3G_F^2}{8\pi} (c_\psi^L-c_\psi^R)^2
\left(\frac{v_{EW}}{\Lambda}\right)^4 \mu^2_{\psi N}(-\Delta^{p,n}_u+\Delta^{p,n}_d+\Delta^{p,n}_s)^2, \nonumber \\
\sigma^{V SI}_{p,n} &=& \frac{32G_F^2}{\pi} c_V^2 \left(\frac{v_{EW}}{\Lambda}\right)^4\mu^2_{\psi N}(2 Y_{u,d}+Y_{d,u})^2.
\label{GaugeOp_dd}
\end{eqnarray}
Similarly, for operators in Eq.~(\ref{HFermion}) we have
\begin{eqnarray}
\sigma^{\phi SI}_{N} &=& \frac{1}{8\pi} \frac{\mu^2_{\phi N} m_N^2 v_{EW}^2}{\Lambda^4} \frac{1}{m_\phi^2} \left(\sum_{q} f_\phi \frac{f_q^{N}}{m_q} \right)^2, \nonumber \\
\sigma^{\psi SI}_{N} &=& \frac{1}{2\pi} \frac{\mu^2_{\psi N}m_N^2 v_{EW}^2}{\Lambda^6}  \left[\left(\sum_{q} f^S_\psi \frac{f_q^{N}}{m_q} \right)^2 + \frac{1}{2}\frac{|\bm{p}|^2 }{m_{\psi}^2} \left(\sum_{q} f^P_\psi \frac{f_q^{N}}{m_q} \right)^2 \right], \nonumber \\
\sigma^{\psi SD}_{N} &=& \frac{6}{\pi} \frac{\mu^2_{\psi N}v_{EW}^2}{\Lambda^6} \left( \sum_q f^T_\psi \delta^N_q \right)^2,  \nonumber \\
\sigma^{V SI}_{N} &=& \frac{1}{2\pi} \frac{\mu^2_{V N} m_N^2 v_{EW}^2}{\Lambda^4} \frac{1}{M_V^2}(f_V)^2 \left(\sum_{q} f_V \frac{f_q^{N}}{m_q} \right)^2. 
\label{HFermion_dd}
\end{eqnarray}
In above equations, $|\bm{p}| \sim $ 1 MeV is the DM momentum in the center of mass frame, $\mu_{\chi N}$ is the DM-nucleon reduced masses (with $\chi= \phi,\psi,V$), and the relevant quark-$Z$ couplings are
$Y_u=\tfrac{1}{2}-\tfrac{4}{3}s_W^2$, and $Y_d=-\tfrac{1}{2}+\tfrac{2}{3}s_W^2$. The parameters $f^{N}_q \equiv m_N^{-1}\left<N\right|m_q\bar{\psi}_q\psi_q\left|N\right>$, $\Delta_{q}^{N}$, and $\delta_{q}^{N}$ indicate the nucleon form factors for scalar, axial-vector, and tensor interactions, respectively. Their values are given by \cite{belanger}
\begin{eqnarray}
\label{nucl-form-fact}
f_u^p &=& 0.023\,,\quad f_d^p = 0.033\,, \quad f_s^p =0.26\,, \nonumber \\
f_u^n &=& 0.018\,,\quad f_d^n = 0.042\,, \quad f_s^n =0.26\,, \nonumber \\
f_{c,b,t}^{p,n} &=& \frac{2}{27}\left(1-\sum_{q=u,d,s} f_q^{p,n}\right), \nonumber \\
\Delta_u^{p,n}&=&0.842\,, \quad \Delta_d^{p,n}=-0.427\,, \quad \Delta_s^{p,n}=-0.085\,, \nonumber \\
\delta_u^{p,n}&=&0.84\,, \quad \delta_d^{p,n}=-0.23\,, \quad \delta_s^{p,n}=-0.05\,.
\end{eqnarray} 

We use XENON100 bounds from Ref. \cite{aprile:2012} for spin-independent (SI) case and Ref. \cite{aprile:2013} for spin-dependent (SD) case to constrain the parameter space given by the relic density. We always use the more constraining choice. 

\section{Analysis of Higgs data}
\label{app:Higgs:fit}

\begin{table}
\caption{The LHC Higgs data used in the analysis, with the Higgs decay channel, production mode, the signal strength normalized to the SM and the correlation coefficient (for details see text). \label{tab:Data-used-in}}
\begin{centering}
\begin{tabular}{cccc}
\hline\hline
Decay channel & Production mode & Signal strength & Correlation \& Reference\tabularnewline
\hline 
\multicolumn{4}{c}{ATLAS}\tabularnewline
\hline
$h\to b\overline{b}$ & VH & $-0.4\pm1.0$ & \cite{key-2}\tabularnewline
\multirow{2}{*}{$h\to ZZ^{*}$} & ggF+ttH & $1.51\pm0.52$ & \multirow{2}{*}{$\rho=-0.5$, \cite{key-1,key-2}}\tabularnewline
 & VBF+VH & $2.0\pm2.1$ & \tabularnewline
\multirow{2}{*}{$h\to WW^{*}$} & ggF+ttH & $0.79\pm0.35$ & \multirow{2}{*}{$\rho=-0.2$, \cite{key-2,key-W}}\tabularnewline
 & VBF+VH & $1.72\pm0.77$ & \tabularnewline
\multirow{2}{*}{$h\to\gamma\gamma$} & ggF+ttH & $1.61\pm0.41$ & \multirow{2}{*}{$\rho=-0.25$, \cite{key-3,key-2}}\tabularnewline
 & VBF+VH & $1.95\pm0.82$ & \tabularnewline
\multirow{2}{*}{$h\to\tau\tau$} & ggF+ttH & $2.3\pm1.6$ & \multirow{2}{*}{$\rho=-0.5$, \cite{key-2}}\tabularnewline
 & VBF+VH & $-0.2\pm1.1$ & \tabularnewline
\multicolumn{2}{c}{$pp\to Zh\to l^{+}l^{-}\textrm{inv}$} & $\textrm{BR}_{\textrm{inv}}=-0.19\pm0.43$ & \cite{ATLAS-inv,Giardino:2012dp}\tabularnewline
\hline 
\multicolumn{4}{c}{CMS}\tabularnewline
\hline 
\multirow{3}{*}{$h\to b\overline{b}$} & VH & $1.0\pm0.5$ & \cite{key-cmsVHb}\tabularnewline
 & VBF & $0.7\pm1.4$ & \cite{key-CMSVBFbb}\tabularnewline
 & ttH & $0.6\pm2.6$ & \cite{Chatrchyan:2013yea}\tabularnewline
\multirow{2}{*}{$h\to WW^{*}$} & ggF+ttH & $0.76\pm0.23$ & \multirow{2}{*}{$\rho=-0.2$, \cite{key-CMS}}\tabularnewline
 & VBF+VH & $0.35\pm0.69$ & \tabularnewline
\multirow{2}{*}{$h\to ZZ^{*}$} & ggF+ttH & $0.90\pm0.45$ & \multirow{2}{*}{$\rho=-0.7$, \cite{key-5,key-CMS}}\tabularnewline
 & VBF+VH & $1.0\pm2.3$ & \tabularnewline
\multirow{2}{*}{$h\to\gamma\gamma$} & ggF+ttH & $0.48\pm0.39$ & \multirow{2}{*}{$\rho=-0.48$, \cite{key-CMS,key-CMSga}}\tabularnewline
 & VBF+VH & $1.70\pm0.88$ & \tabularnewline
\multirow{2}{*}{$h\to\tau\tau$} & ggF+ttH & $0.68\pm0.80$ & \multirow{2}{*}{$\rho=-0.46$, \cite{key-CMS}}\tabularnewline
 & VBF+VH & $1.61\pm0.83$ & \tabularnewline
\hline\hline 
\end{tabular}
\par\end{centering}
\end{table}

In our fitting procedures we follow the method adopted in references~\cite{Cacciapaglia:2012wb,Belanger:2012gc,Fajfer:2013wca,Dumont:2013wma}. The
latest available LHC Higgs data are presented in Table ~\ref{tab:Data-used-in}.
Measurements are reported in terms of signal strengths normalized
to the SM predictions
\begin{equation}
\mu_{(k)}^{i}=\frac{\sigma_{(k)}}{\sigma_{(k)}^{SM}}\frac{\mathcal{B}_{i}}{\mathcal{B}_{i}^{SM}}\,,
\end{equation}
where index $i$ represents the decay mode, while $k$ denotes different production
channels. ATLAS and CMS also combine different production sub-channels
for a given decay mode to provide separation into production mechanisms.
Results are presented in 2D plots in which gluon-gluon fusion (ggF)
and associated production with a top pair (ttH) are combined as one
signal ($\mu_{(ggF+ttH)}$), while vector boson fusion (VBF) and associated
production with a gauge boson (VH) as another, ($\mu_{(VBF+VH)}$).
In this case, we parametrize the likelihood with
\begin{equation}
\chi_{1}^{2}=\underset{i}{\sum}\left(\begin{array}{c}
\mu_{(ggF+ttH)}^{i}-\hat{\mu}_{(ggF+ttH)}^{i}\\
\mu_{(VBF+VH)}^{i}-\hat{\mu}_{(VBF+VH)}^{i}
\end{array}\right)^{T}V_{i}^{-1}\left(\begin{array}{c}
\mu_{(ggF+ttH)}^{i}-\hat{\mu}_{(ggF+ttH)}^{i}\\
\mu_{(VBF+VH)}^{i}-\hat{\mu}_{(VBF+VH)}^{i}
\end{array}\right),
\end{equation}
where the correlation matrices are given by
\begin{equation}
V_{i}=\left(\begin{array}{cc}
\left(\hat{\sigma}_{(ggF+ttH)}^{i}\right)^{2} & \rho^{i}\hat{\sigma}_{(ggF+ttH)}^{i}\hat{\sigma}_{(VBF+VH)}^{i}\\
\rho^{i}\hat{\sigma}_{(ggF+ttH)}^{i}\hat{\sigma}_{(VBF+VH)}^{i} & \left(\hat{\sigma}_{(VBF+VH)}^{i}\right)^{2}
\end{array}\right).
\end{equation}
Best-fit values ($\hat{\mu}$), variances ($\hat{\sigma}$) and correlations
($\rho$) are obtained from the plots provided by the experiments and listed in Table \ref{tab:Data-used-in}. 

Other data are given in terms of signal strengths with specified production mechanism. In this case, we parametrize the likelihood with
\begin{equation}
\chi_{2}^{2}=\underset{i}{\sum}\left(\frac{\mu_{i}-\hat{\mu}_{i}}{\hat{\sigma}_{i}}\right)^{2}.
\end{equation}
The total $\chi^{2}$ function is given by the sum of all the contributions. In order to confront the DM model to the data, we express all signal strengths ($\mu$) in terms of model parameters and minimize
$\chi^{2}$ to find the best fit point. The best fit regions are defined by appropriate cumulative distribution functions.


\end{document}